%% file: article_jcp_v2.tex
\documentclass[a4paper]{article}

\usepackage[margin=1in]{geometry}
\usepackage[utf8]{inputenc}
\usepackage[T1]{fontenc}
\usepackage{lmodern}
\usepackage[english]{babel}

\usepackage{amsmath, amsthm, amsfonts, amssymb, mathtools, xparse}
\usepackage{cite}
\usepackage{graphicx, wrapfig, float}
\usepackage{enumitem}
\usepackage{authblk}
\usepackage{hyperref, hypcap}
\usepackage{xcolor, color, soul}

\usepackage[draft, textsize=tiny]{todonotes}   
\presetkeys{todonotes}{color=yellow!30, linecolor=black}{}

\hypersetup{colorlinks=true,allcolors=blue}
\theoremstyle{plain}
\newtheorem{thm}{Theorem}[section] 

\theoremstyle{definition}
\newtheorem{defn}[thm]{Definition} 

\DeclarePairedDelimiter{\abs}{\lvert}{\rvert}
\DeclarePairedDelimiter{\norm}{\lVert}{\rVert}

\begin{document}

\title{Computational Fluid Dynamics on 3D Point Set Surfaces}

\author[1]{Hassan Bouchiba}
\author[2]{Simon Santoso}
\author[1]{Jean-Emmanuel Deschaud}
\author[2]{Luisa Rocha-Da-Silva}
\author[1]{François Goulette}
\author[2]{Thierry Coupez}

\affil[1]{MINES ParisTech, PSL Research University, Centre for robotics, Paris, France}
\affil[2]{Ecole Centrale Nantes, High Performance Computing Institute, Nantes, France}

\renewcommand\Affilfont{\itshape\small}
\renewcommand\Authands{ and }

\maketitle

\begin{abstract}
  \input{abstract}
\end{abstract}

\begin{flushleft}
\end{flushleft}

\input{1_introduction}
\input{2_related_work}
\input{3_point_cloud_acquisition_and_processing_reduc}

\input{4_implicit_surface_definition}

\input{5_anisotropic_mesh_apadtation}

\input{6_results}
\input{7_conclusion}
\input{aknowledgement}

\clearpage
\bibliography{article_jcp_v2}


\end{document}

%% file: abstract.tex
Computational fluid dynamics (CFD) in many cases requires designing 3D models manually, which is a tedious task that requires specific skills. In this paper, we present a novel method for performing CFD directly on scanned 3D point clouds. The proposed method builds an anisotropic volumetric tetrahedral mesh adapted around a point-sampled surface, without an explicit surface reconstruction step. The surface is represented by a new extended implicit moving least squares (EIMLS) scalar representation that extends the definition of the function to the entire computational domain, which makes it possible for use in immersed boundary flow simulations. The workflow we present allows us to compute flows around point-sampled geometries automatically. It also gives a better control of the precision around the surface with a limited number of computational nodes, which is a critical issue in CFD.

%% file: 1_introduction.tex
\section{Introduction}
\label{sec:introduction}

\subsection{General framework}

In many cases, numerical simulation of physical phenomena such as external aerodynamics, acoustics wave propagation, or heat transfer, requires a 3D model of the object. In a typical conception framework usually, the CAD model of the object exists and thus, can be used for the numerical simulation. However, when it comes to studying existing objects, such as for air ventilation assessment or retro-engineering the object should be manually modeled, and 3D modeling is a time-consuming task that requires high skills.

For the past 30 years, 3D scanning techniques have been growing. In particular, for urban environments mobile mapping \cite{goulette_integrated_2006} enables fast and accurate acquisition with vehicles driven at regular speed in traffic. These techniques are an interesting alternative to manual modeling as they produce accurate high-resolution models. Unfortunately, 3D scanning techniques never directly yield a watertight and manifold surface mesh of the scanned object or scene. Instead, they generate a noisy and unevenly spaced set of points sampled on the object's surface, which is called a point cloud. Although it can be very dense, it does not hold any topological or connectivity information which makes it impossible to use as is, in typical numerical simulation frameworks. In addition, these models are usually massive as today's scanners produce up to one million points per second.


\begin{figure}
  \centering
  \includegraphics[width=1.0\textwidth]{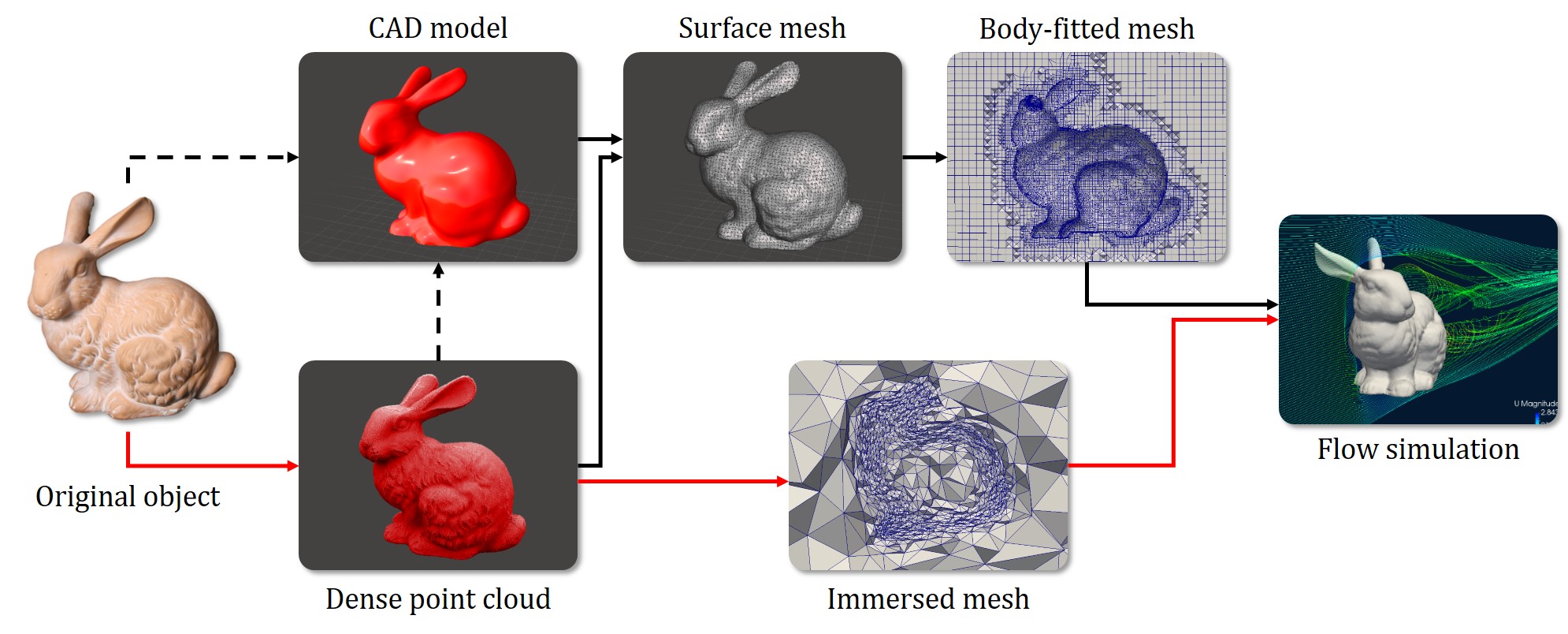}
  \caption{Different workflows for simulating flows around real-life objects. Dashed arrows represent manual steps, and plain arrows represent automatic or semi-automatic steps. Black arrows outline the \textit{classic workflows}. The proposed method is depicted with the red arrows. It involves less user interaction as the simulation grid is directly computed from the original point cloud.}
  \label{fig:contrib}
\end{figure}

The main goal of this article is to propose a novel method for bringing point clouds directly into numerical simulation frameworks to automate the manual 3D modeling step, and then open up a new field of simulation cases on real-life complex objects and scenes. Although we believe that the proposed method can be generalized to a wide range of applications, we will only focus on finite element incompressible unsteady flow simulation. Throughout the rest of the article, we will refer to it by the more general acronym CFD, for computational fluid dynamics.

Figure~\ref{fig:contrib} illustrates different ways to simulate flows around an existing object. As written above the first alternative is manual or semi-automatic CAD modeling from the point cloud. Examples of this approach are given in Section~\ref{sec:related_work}. It is user involving and the CAD model is not as detailed as the original scan as CAD is built upon simple 3D primitives or defined in a parametric space. This makes this alternative suited only for structured environments, such as buildings.

As depicted in Figure~\ref{fig:contrib}, another alternative to traditional and tedious CAD modeling is building an explicit surface mesh from a dense point cloud, and then building a body-fitted mesh upon it. Although many efforts have been made in the computational geometry community to make a surface mesh from a point cloud \cite{hoppe_surface_1992, amenta_power_2001, kazhdan_poisson_2006}, we believe that bringing points \textit{more directly} in the simulation is a more promising method for the following reasons:
\begin{itemize}
  \item An explicit surface mesh is a choice for the surface representation. Once it is built, we make a strong assumption about the topology of the object. Whereas the point cloud is the closest data from the measurement as it is the direct output of the scan process. In addition, for complex objects, such as outdoor scenes, surface reconstruction methods fail to reconstruct an explicit and coherent mesh.
  \item Traditional body-fitted CFD grids have very specific requirements for the size and shape of the cells on the boundary of the object. If such a mesh is built upon a surface mesh, these requirements must be inferred in the surface mesh building process, too, which is possible when the mesh comes from a CAD model but much harder when the mesh is built from a point cloud.
  \item Surface reconstruction methods applied to complex geometries produce over-detailed models with millions of triangles, which is not in line with the current computational power available for CFD. Although making different levels of detail of a mesh is possible, point clouds intrinsically carry multi-resolution information.
\end{itemize}


Our contribution is to present a method for simulating flows around 3D point clouds. This contribution is a new step toward automated numerical simulation. To the authors' knowledge, this work is the first of its kind.

This contribution relies on two points. The first one is that the only explicit representation used for the surface geometry is a 3D point cloud. As depicted in Figure~\ref{fig:contrib}, using only the points involves less work for the final user and fewer intermediate steps while making the most of the geometric precision that a dense point cloud can provide.

The second point is a novel implicit surface representation called EIMLS, for extended implicit moving least squares, to define an implicit surface from a point cloud. Unlike state-of-the-art implicit representations, this new representation allows to define the implicit function far from the surface, which is not usually needed for explicit surface reconstruction but is of great importance for immersed boundary simulation.

\subsection{The proposed method}

\begin{figure}
  \centering
  \includegraphics[width=0.6\textwidth]{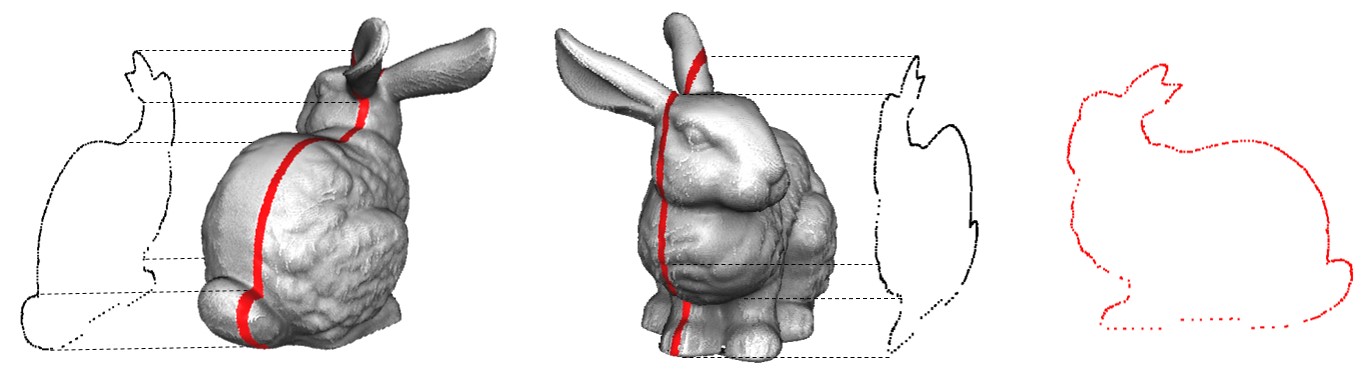}
  \caption{Slice of the Stanford Bunny used as a 2D dataset in this article. The slice contains 557 2D points with normals. The normals have been computed in 3D and then projected in 2D. See Section~\ref{sec:point_cloud_acquisition_and_processing} for more details on the 3D dataset and related preprocessing.}
  \label{fig:dataset_2d}
\end{figure}

\begin{figure}
  \centering
  \includegraphics[width=\textwidth]{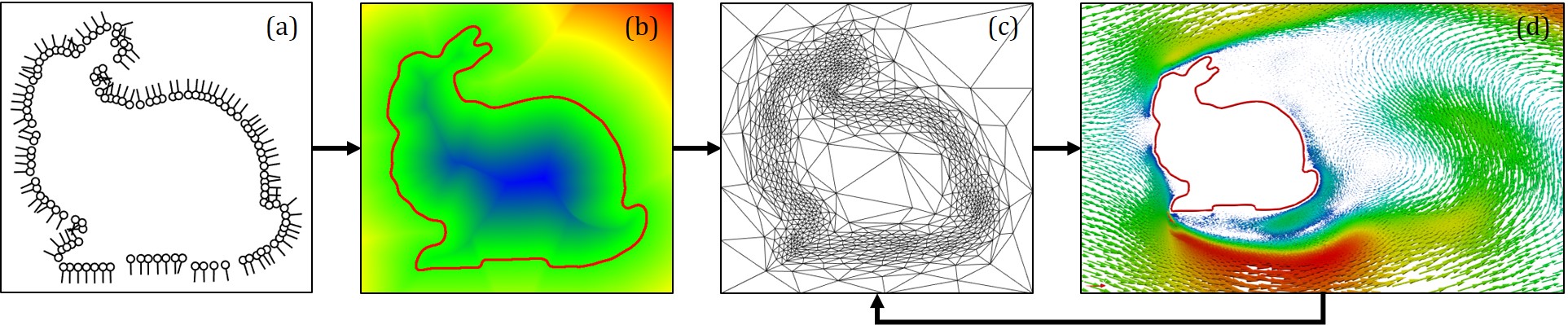}
  \caption{Overview of our workflow. (a) Point cloud with oriented normals. (b) Implicit scalar function computed by the EIMLS. Its iso-zero level set is depicted in red. (c) Anisotropic mesh adapted around the surface sampled by the point cloud. (d) Flow computed with the implicit boundary and finite element method.}
  \label{fig:our_method}
\end{figure}

Figure~\ref{fig:dataset_2d} shows the 2D dataset we use to illustrate the proposed method. It has been obtained by slicing the well-known Stanford Bunny 3D point cloud model. This dataset contains 2D points with unit 2D normal vector. Although the model is 2D, we show in Section~\ref{sec:implicit_surface_definition} that the model is representative of a \textit{real-world} dataset.

The proposed method is depicted in Figure~\ref{fig:our_method}. First, a $d$-dimensional ($d=2$ or $3$) point set with normals (a) is used to compute a signed scalar function (b) that represents implicitly the surface of the model sampled by the point cloud. In Section~\ref{sec:point_cloud_acquisition_and_processing}, we discuss how a 3D point cloud can be acquired and preprocessed to get rid of some artifacts and to obtain normals. Then in Section~\ref{sec:implicit_surface_definition} we introduce the proposed EIMLS method for computing the implicit scalar function from the point cloud. An immersed anisotropic unstructured grid (c) can then be built and adapted around the surface of the model which is the zero level set of the implicit function. In Section~\ref{sec:anisotropic_mesh_adaptation}, we show how to perform anisotropic mesh adaptation on point-sampled geometry. The mesh is then used as space discretization for an immersed boundary transient incompressible Navier-Stokes finite element solver to simulate the behavior of a flow around the surface of the object (d). Then, during the simulation, the mesh is refined around both the surface and the flow thanks to \textit{a posteriori} error metric. The implicit scalar function is recomputed only at each remeshing step at the nodes of the grid. Then its interpolation over the mesh is used during the flow computation steps. Numerical results for real-world datasets are presented in Section~\ref{sec:results}.

%% file: 2_related_work.tex
\section{Related work}
\label{sec:related_work}

Our work is at the crossroads of multiple scientific domains: immersed boundary computational fluid dynamics, geometry processing and anisotropic mesh adaptation. Thus, it is hard to produce a comprehensive overview of the related work to the proposed method. A more detailed overview of related work is given in each specific part of this article.

First, we can compare the proposed work to the idea of using points as primitives for numerical simulation. This idea has been extensively used in the context of animation. \cite{gross_point-based_2011} give a general overview of point-based animation. In this context, \cite{muller_position_2007} introduced the famous position-based dynamics (PBD) that replaced the traditional force-based simulation framework with a position-based one. Acting directly on positions lets the simulation be more stable in the context of real-time simulation. This original work has been extensively used in the physics-based animation community. More recently \cite{macklin_unified_2014} proposed a unified solver for rigid and deformable bodies, cloths, and fluids with impressive real-time results. Although these methods are point-based, they remain far from the proposed work as the points are used here as support for the simulation and not to define its boundary.

Finally, the closest idea to the proposed method is performing numerical simulation on real-world data. This idea has been exploited in the context of gaming and user interaction. \cite{newcombe_kinectfusion:_2011, izadi_kinectfusion:_2011} demonstrated an interactive simulation of rigid bodies interacting with a scene live scanned by a commodity depth sensor. The interesting part of this work is that no explicit representation of the scene is stored internally as it is represented as a signed scalar function that is updated in real time. However, as the original points are not kept, the maximum resolution of the 3D model is fixed by the regular grid used to store the points. The underlying surface representation used is the one introduced by \cite{curless_volumetric_1996}. A more detailed comparison with the proposed surface representation is given in Section~\ref{sec:implicit_surface_definition}.

In the context of culture heritage preservation, \cite{oreni_beyond_2014, barazzetti_bim_2015} proposed a semi-automatic method based on commodity CAD software to build a building information modeling (BIM) model from 3D laser scanned historical buildings. They then used  the reconstructed CAD model to perform finite element analysis (FEA). Although their method can handle massive datasets, it is not fully automatic and requires specific skills with CAD software. \cite{castellazzi_laser_2015} proposed an interesting semi-automatic workflow to perform FEA on point clouds. They cut their 3D model into 2D point cloud slices that are manually cleaned and processed. The slices are then stacked in 3D to produce a voxelized 3D model which is used for FEA simulation. However, the voxelization procedure is limited as the maximum resolution of the model is fixed by the voxel size, which is sufficient for global structural behavior simulation but not sufficient for fine fluid simulations near the surface.

Recent work in the domain of medical simulation have proven the ability to simulate mechanical behavior or fluid flows on organs from CT scans. \cite{venkatasubramaniam_comparative_2004} presented an automatic method for generating FEA models from CT scans to study abdominal aortic aneurysms. More recently, \cite{chnafa_elucidating_2013} presented a mesh-based method for simulating blood flow during the heart cycles. Large eddy simulation (LES) is used on a deformable mesh that has been registered to dynamic CT scans. By simulating the heart behavior based on real CT scans, the simulation is patient-specific which opens up many interesting applications. Nevertheless, this method is based on an explicit mesh that is not well suited for CFD on complex geometries. \cite{mihalef_patient-specific_2011} also presented a patient-specific hemodynamics (dynamics of blood flows) simulation from CT scans, but instead of using an explicit surface mesh as the input for the simulation, they used an immersed boundary flow solver and represented the heart with a level set computed from the CT scans. This approach is the closest to the method proposed in this paper, but the level set computation was made easier as the input data was structured and volumetric whereas the input data we propose to study in this article is unstructured and sampled on surfaces.

%% file: 3_point_cloud_acquisition_and_processing_reduc.tex
\section{Point cloud acquisition and processing}
\label{sec:point_cloud_acquisition_and_processing}

\subsection{Datasets}

\begin{figure}
  \centering
  \includegraphics[width=.8\textwidth]{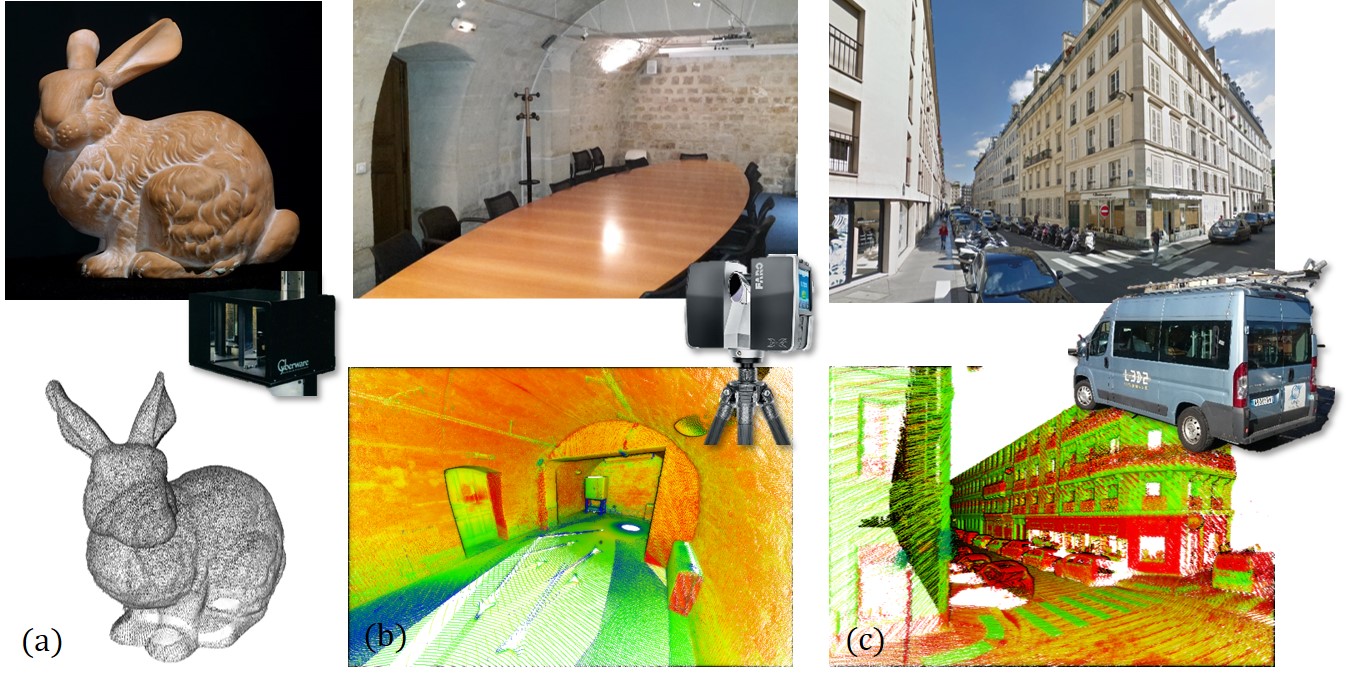}
  \caption{The three 3D datasets used in this article. (a) Scan of an object: \textbf{bunny} dataset. (b) Terrestrial Laser Scan of a meeting room: \textbf{room} dataset. (c) Mobile Mapping Scan: \textbf{street} dataset. (b) and (c) are colored by the lidar remittance which is not used in the proposed method.}
  \label{fig:datasets_3d}
\end{figure}

Three-dimensional scanning techniques can be used to make 3D models from real-life objects. Today, these techniques are widely used for industrial and consumer applications. The 3D measurement can be performed thanks to various physical phenomena: Contact scanners use the resistive force of a surface, laser scanners use the Lambertian reflectivity of a surface to measure the time of flight of a laser beam shot hundreds of thousands times per second, and CT scans and other related volumetric scanners can produce 3D slices of an object. In this work, we focus on surface 3D scanning techniques. \cite{zhao_direct_2016} present the same adaptive finite element simulation framework on volumetric data.

To illustrate our work, we chose three different point cloud datasets depicted in Figure~\ref{fig:datasets_3d}:
\begin{enumerate}[label=(\alph*)]
  \item \textbf{bunny}: Introduced in the previous sections of this article, this dataset has been acquired by the Stanford University Computer Graphics Laboratory in 1993-4. The model has been scanned thanks to a Cyberware 3030 MS scanner from a clay statue. This well-known dataset is made up of 10 range images registered thanks to a modified ICP algorithm \cite{turk_zippered_1994}. It is 15~cm tall and has 362,270 points. Behind its apparent geometric simplicity, this model has several artifacts as outliers and holes in its bottom part (see Section \ref{sec:implicit_surface_definition} for more details). Its surface also presents interesting sharp details.

  \item \textbf{room}: Saint Jacques is a meeting room in the main building of Mines ParisTech school in Paris. It is 7 m by 10 m by 4 m wide, has 13 million points, and has ancient brick walls and a dome ceiling. It has been acquired thanks to a Faro Focus X130 3D Terrestrial Laser Scanner (TLS). Five scans have been acquired in various positions in the room. The center table that can be seen in the photo has been manually removed from the scan for convenience, as it has not been totally scanned. The mean local accuracy of each scan is 2 mm, but the scan contains many outliers, mainly due to scanner beams hitting sharp corners.

  \item \textbf{street}: The Madame street is located in the 6th district in Paris. It is a publicly available dataset for point cloud segmentation and classification \cite{serna_paris-rue-madame_2014}. The dataset has been acquired with the L3D2 prototype of the CAOR robotics laboratory at Mines ParisTech school. It is a Mobile Mapping System (MMS) \cite{goulette_integrated_2006} equipped with an INS/GPS localization system and a Velodyne HDL32E lidar. This latter has an accuracy of 2~cm, and its 32 laser beams, mounted on a 10 Hz spinning head, produce up to 700,000 points per second. An MMS is designed to acquire urban 3D data with a high productivity rate, as the vehicle moves at normal speed in traffic. The scan used here is a 30~m portion of the whole dataset composed of 2 million points. It is the most challenging dataset as MMS produces more noisy point clouds (5~cm mean error) due to cumulated errors on the localization sensors and the lidar. As can be seen in Figure~\ref{fig:datasets_3d}, the dataset has also a lot of hidden parts.
\end{enumerate}

These datasets are chosen to cover a wide range of 3D scanning techniques. This decision is also motivated by specific CFD applications, trying to cover a broad range of industrial uses. Each simulation test case is described in Section~\ref{sec:results}.

\subsection{Point cloud preprocessing}

In Section \ref{sec:implicit_surface_definition}, we present the EIMLS implicit scalar function computed from point clouds. To do so, the point cloud must be preprocessed to remove the outliers, be subsampled, and compute surface normals. We present these preprocessing steps on the three datasets. More details on why these specific preprocessing steps are needed are given in Section \ref{sec:implicit_surface_definition}.

\subsubsection*{Outliers removal}
Real-world point clouds always contain outlier points due to laser echo misinterpreted by the scanner software. An easy approach for removing these outliers is to compute the local density on each point and suppress points with too low density. The density can be computed from various ways. For the MMS dataset, we chose to compute the distance to the third closest point and to use a 30~cm threshold.

The TLS scan features more challenging structured noise due to the laser hitting sharp corners and producing a ghost trailing set of points in the laser shooting axis. To get rid of these points, we compute the angle between the point normal and the direction between the point and the scanner center. The points seen with a grazing angle (typically $2^{\circ}$) are removed.

\subsubsection*{Subsampling}
The TLS and MMS scans present anisotropy due to the difference between the angular resolutions of the sensor in its various directions. For example, the Velodyne HDL32E has a vertical angle resolution (the angle between each consecutive laser beam) of $1.33^{\circ}$ and a horizontal resolution of $0.1^{\circ}$. This anisotropy should be taken into account in the later used algorithms by, for exampl, using Hough accumulators \cite{boulch_fast_2012} for normals computing. We use a simpler approach by subsampling the \textbf{room} and \textbf{street} datasets on a space criterion. To do so, we built an octree data structure \cite{elseberg_one_2013} with a fixed leaf size criterion. In practice, we used 2~cm.

\subsubsection*{Normals computing}
\cite{hoppe_surface_1992} introduced a simple method for computing normals over a point cloud. The normal is defined as the eigenvector associated with the smallest eigenvalue of the covariance matrix of each point neighborhood. This simple formulation with a neighborhood of 100 points gives sufficient results for the three datasets. For the orientation of the normals we used the scanner origin information. Nonetheless, computing normals for a point cloud can be a challenging problem, and more refined methods can produce better results for more complex geometries \cite{boulch_fast_2012, boulch_deep_2016}.

%% file: 4_implicit_surface_definition.tex
\section{Implicit surface definition}
\label{sec:implicit_surface_definition}

\begin{table}
  \centering
  \begin{tabular}{ll}
    \hline
    Notations 1 & Definition \\
    \hline
    $d$                                               & dimension of space \\
    $\mathcal{N} \subset \mathbb{N}$                  & set of point indices \\
    $\mathcal{P} = \{ (p_i, n_i), i\in\mathcal{N} \}$ & oriented point cloud: points and normals \\
    $p_i \in \mathbb{R}^d$                            & $d$-dimensional position of the $i$-th point \\
    $n_i \in \mathbb{S}^{d-1}$                        & unit normal of the $i$-th point (lying on the unit sphere)\\
    $\mathcal{S}_\mathcal{P}$                         & underlying smooth surface sampled by $\mathcal{P}$ \\
    $\alpha:\mathbb{R}^d \rightarrow \mathbb{R}$      & implicit function that represents $\mathcal{S}_\mathcal{P}$\\
    \hline
  \end{tabular}
  \caption{Notations and definitions for Section~\ref{sec:implicit_surface_definition}}
  \label{tab:implicit_notations}
\end{table}

In this section, we show, given an oriented point set $\mathcal{P}$, how to design a scalar function $\alpha:\mathbb{R}^d \rightarrow \mathbb{R}$ that defines implicitly the underlying smooth surface $\mathcal{S}_\mathcal{P}$ sampled by the point cloud. Near the surface, the implicit function should behave as a distance function to $\mathcal{S}_\mathcal{P}$. The reader can refer to Table~\ref{tab:implicit_notations} for the notations used in this section.

After we present related work on surface reconstruction (subsection~\ref{subsec:implicit_surface_definition:related_work}), we present a new implicit surface representation suitable for defining the boundary conditions of an immersed boundary numerical simulation (subsection~\ref{subsec:implicit_surface_definition:eimls}). In particular, we describe the constraints related to this specific application framework.

\subsection{Related work}
\label{subsec:implicit_surface_definition:related_work}

A surface is a $(d-1)$ manifold of $\mathbb{R}^d$; it can be represented in many ways. We usually differentiate the explicit representations from the implicit ones. Explicit representations describe a surface with a discrete set. For instance, in 3D a surface can be described with a set of triangles to form a triangle mesh, or a set of points to form a point cloud. Implicit representations describe a surface indirectly, for example, with the level set of a scalar function or the fixed point of a projection operator. 

Passing from a point cloud to any other surface representation is called \textit{surface reconstruction}. It is an important research topic in computer graphics and computational geometry. For an exhaustive and recent study, the reader can refer to \cite{berger_state_2014}.

\paragraph*{Explicit surface reconstruction}
There exist methods for building a surface mesh directly from a point cloud. These methods are usually called explicit methods. The \textit{ball-pivoting algorithm (BPA)} \cite{bernardini_ball-pivoting_1999} interpolates the points by rolling a virtual ball between them. Each time the ball touches three points, a new triangle is created. This method and its variants are heuristic-based and fail when the local curvature is higher than the ball radius. Another kind of methods are those based on Vorono\"{i} diagrams, such as the crust \cite{amenta_surface_1999} and the power crust algorithm \cite{amenta_power_2001}. They are essentially based on the property that the surface mesh is a subset of the dual of the Vorono\"{i} diagram of the point cloud. The main advantage of these methods is that they provide good theoretical guarantees on the method itself and on the output mesh quality. A good overview of Vorono\"{i}-based surface reconstruction methods is given by \cite{cazals_delaunay_2006}.

Usually, explicit methods are interpolatory: They try to link all the points to produce a surface mesh. They are then not resilient to massive and noisy point clouds, and thus, not compatible with real-life complex point clouds. As prior mentioned, we believe that building a surface mesh from a point cloud is making a strong assumption on the topological and geometric properties of the surface, and then only replaces an explicit representation by another. Therefore, we find it more convenient to build an implicit representation based directly on the point cloud instead.

\paragraph*{Implicit surface reconstruction}
Implicit methods build an indirect representation of a surface. Literature on implicit surface reconstruction abounds. Curless and Levoy introduced a method for building a truncated signed distance function (TSDF) from range measurements \cite{curless_volumetric_1996}. Each depth frame is used to carve a regular grid, labeling not only the space near the measured surface but also all the visited space along the ray between the measurement and the sensor, thus providing important additional topological information. This method has recently received renewed interest. KinectFusion \cite{newcombe_kinectfusion:_2011} uses this method to register in real time depth images from a low-cost depth camera and build a surface from it. Kazdhan et al. \cite{kazhdan_poisson_2006, kazhdan_screened_2013} recently showed that the surface reconstruction problem can be formulated as a Poisson equation. They solve it thanks to a finite element solver on an octree grid. \textit{Poisson reconstruction} is based on the use of oriented normals. Orienting normals can be difficult; therefore, other methods focus on using unoriented normals. Alliez et al. \cite{alliez_voronoi-based_2007} showed that surface reconstruction from unoriented normals can be expressed as a generalized eigenvalue problem. Mullen et al. \cite{mullen_signing_2010} tackled the problem differently by first computing a robust unsigned distance field and then statistically signing it.

\paragraph*{Explicit surface reconstruction from implicit representation}
Nonetheless, care must be taken in the proposed classification of surface reconstruction methods, as their output is usually a surface mesh, although the underlying method is implicit. The mesh is built only in the final step, for example, for visualization purposes. There exists methods for generating a surface mesh from the level set of an implicit scalar function. Lorensen and Cline first introduced the \textit{marching cubes algorithm} \cite{lorensen_marching_1987} that extracts a triangle mesh from the level set of a scalar function sampled on a regular grid. Treece et al. \cite{treece_regularised_1998} extended this method to better handle ambiguities on the reconstructed surface of the original method by sampling the function on a tetrahedral grid. \cite{schaefer_dual_2004} introduced a variant of the original method to extract a mesh from an octree: dual marching cubes. This method produces a mesh with adaptive-sized triangle patches to limit the over-subdivision of the surface produced by the original method.

\paragraph*{Implicit surface reconstruction by moving least squares}
These implicit methods are \textit{global} which means that the implicit scalar function is computed on a discrete grid (regular, octree, or unstructured), and the entire point set is used in the process. There exists also another set of methods that are qualified as \textit{local}: The implicit function is defined on any point of the $d$-dimensional space, and its computation involves only on a small subset of the whole point cloud.

Among these, methods based on moving least squares (MLS) provide a good theoretical background at small computational cost. The original work of Hoppe et al. \cite{hoppe_surface_1992} is the first local implicit signed scalar function definition from an oriented point cloud. However, the first MLS surface definition is derived from the original work of Levin \cite{levin_mesh-independent_2004}. He introduced a projection operator computed in two steps: First, given a query point $q\in\mathbb{R}^d$, a plane is fitted on a neighborhood of $q$ in the point cloud. A higher-order polynomial function is then fitted locally, and $q$ is projected on the approximated surface. The surface is then defined as the stationary points of this projection operator. \cite{alexa_computing_2003} then used this definition to render surfaces. Surfaces defined by a local projection operator are usually referred to as point set surfaces (PSS).

Shen et al. \cite{shen_interpolating_2005} introduced a MLS-based definition of an implicit function to estimate a surface from a polygon soup (i.e., a surface mesh without connectivity). Kolluri \cite{kolluri_provably_2005} then applied this definition to point clouds. He also proved some theoretical properties of the method. This implicit definition is usually referred to as implicit moving least squares (IMLS).

PSS and IMLS formulations have intrinsic limitations when the input point cloud quality is poor. Fleishman et al. \cite{fleishman_robust_2005} first used robust statistics to recover a piecewise smooth surface from a modified projection operator. \"{O}ztirelli et al. \cite{oztireli_feature_2009} also used robust statistics to provide a robust definition of IMLS called \textit{RIMLS}. These methods help better recover sharp features in noisy point clouds. Guennebaud and Gross \cite{guennebaud_algebraic_2007} extended the projection and the implicit definitions with the algebraic point set surfaces (APSS) algorithm. They fit algebraic spheres to better handle low-sampled and sharp point clouds.

In addition, all these methods include an averaging process that is resilient to noise in the point cloud.

\subsection{Designing an implicit scalar function}
\label{subsec:implicit_surface_definition:eimls}

In this subsection, we explain the reasoning that led us to define our new extended implicit surface representation: EIMLS. This reasoning is twofold. First, studying the constraints imposed by dealing with real-world point clouds and those imposed by the rest of our pipeline, we show that the best-suited state-of-the-art surface representation is IMLS. Then, after presenting it, we show why it needs to be extended to be used in our application framework.

\subsubsection{Design constraints}

\begin{figure}
  \centering
  \includegraphics[width=0.5\textwidth]{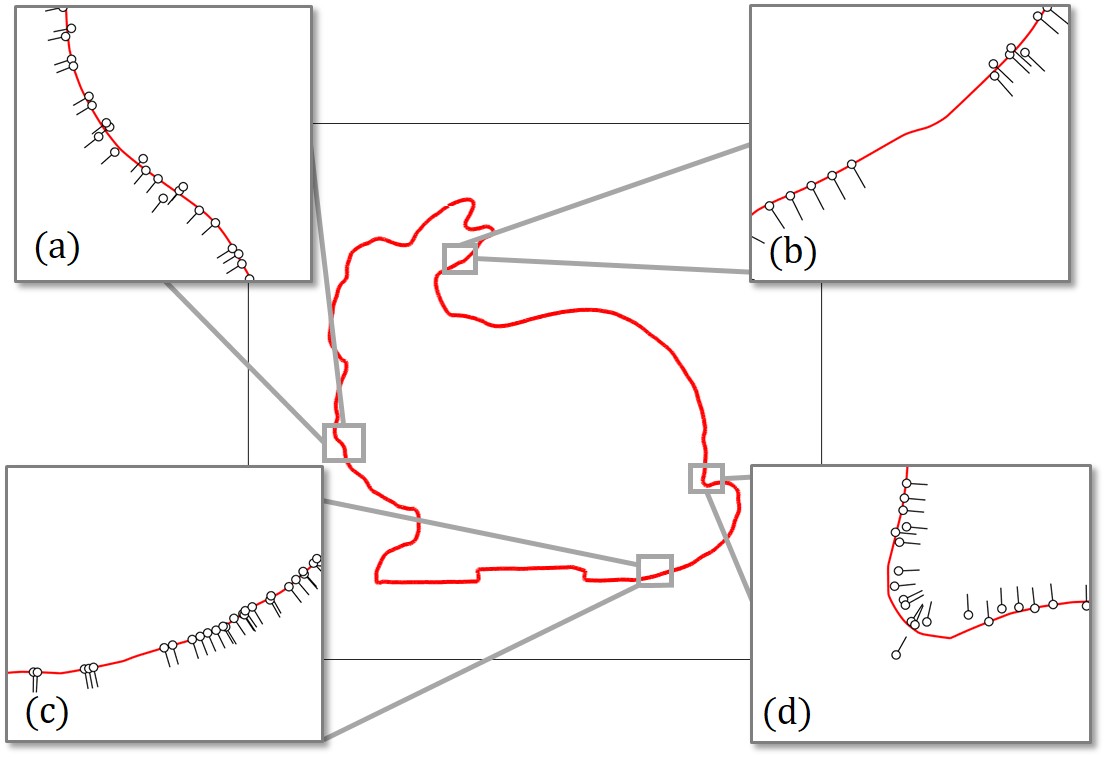}
  \caption{Main artifacts that can be found in point clouds illustrated on a 2D model. (a) Noise. (b) Missing data. (c) Non-uniform sampling. (d) Outliers.} 
  \label{fig:pc_artifacts}
\end{figure}

\paragraph*{Real-life point cloud artifacts}
We represent the usual artifacts \cite{berger_state_2014} that can be found on real-life point clouds on the 2D dataset in Figure~\ref{fig:pc_artifacts}: noise, holes, non-uniform sampling, and outliers. We already dealt with outliers in Section~\ref{sec:point_cloud_acquisition_and_processing} by removing them with appropriate preprocessing. Thus, the proposed implicit surface representation should deal with the three first ones. They represent the data-related design constraints.

\paragraph*{Signed or unsigned?}
Another question that we should address is which kind of implicit representation we want to define the surface. We saw that an implicit surface can be defined from a point cloud by either a projection operator or a scalar function. Projection operators usually involve iterative local optimization which is more computationally intensive than computing a simple scalar function. Therefore, we prefer the scalar function.

We then wonder whether the scalar function must be signed or unsigned. If it is signed, the surface can be retrieved from the scalar function by extracting a level set (typically the iso-zero). Unsigned scalar functions are easier to build, as they do not need the normals to be oriented. Unsigned scalar functions involve local minimum research to extract the surface. However, it should be noted that the flow solver does not need the implicit function to be signed to enforce the boundary conditions.

Nonetheless, signed implicit functions have a major benefit compared to unsigned ones: Signed functions present additional topological information about the interior and the exterior of the surface. This additional property usually allows to close holes in the point cloud models, addressing the problem of missing data. In addition, this interior/exterior information can  later be used in the numerical simulation framework to specify physical properties on the object. Thus, although unsigned scalar functions are easier to build, we need a signed one.

\paragraph*{Global or local?}
Should the implicit function be computed once on a predefined grid (global method) or at execution time (local method)? Global methods are more robust to point cloud artifacts but are more computationally intensive and less versatile than local methods. Local methods need only a neighborhood of the point cloud to be computed at a given point of the space. They are then better suited to an adaptive meshing context as we use in the proposed framework, as the grid is constantly modified at each iteration of the simulation.

\subsubsection{Implicit moving least squares (IMLS)}

\begin{figure}
  \centering
  \includegraphics[width=0.7\textwidth]{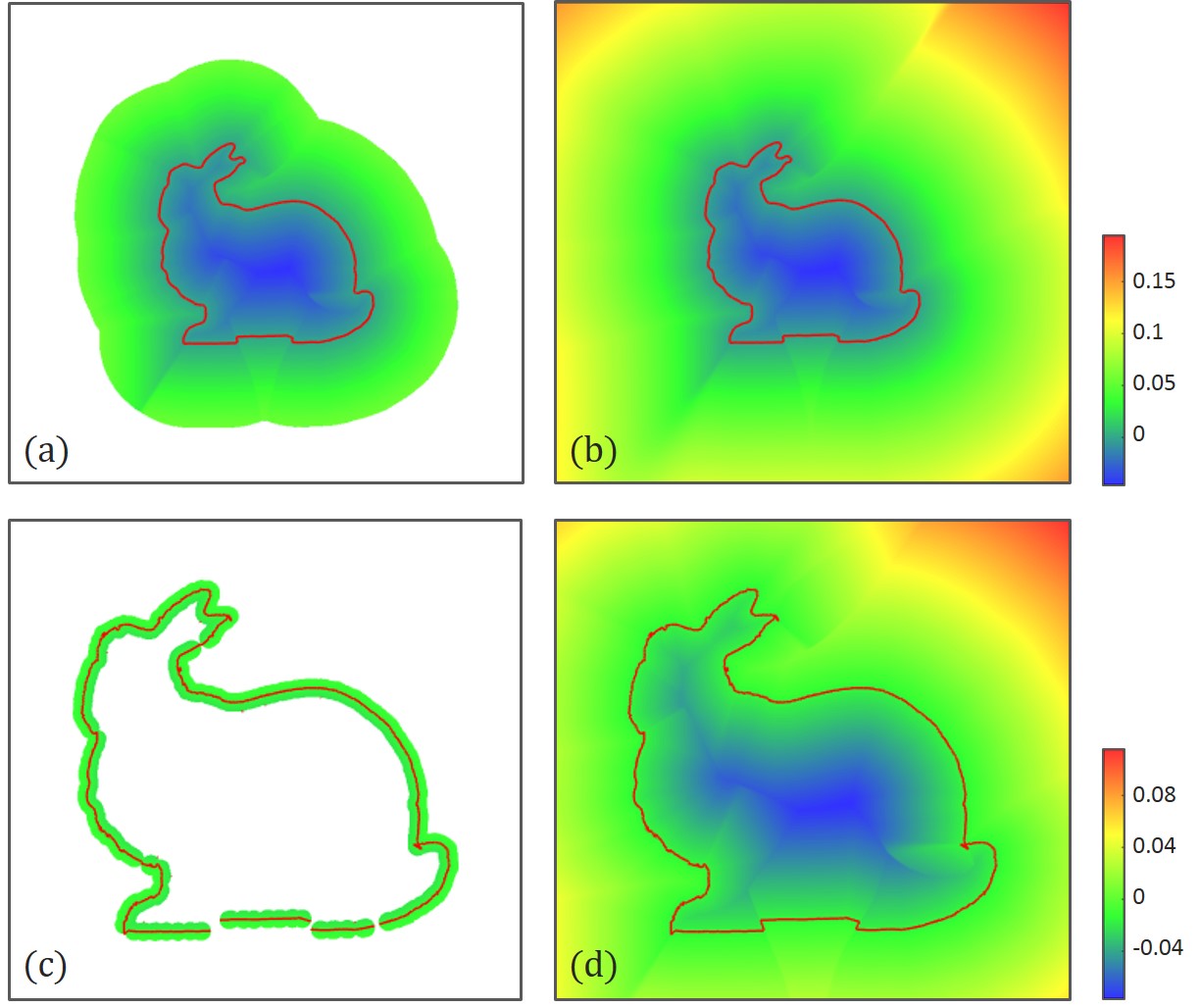}
  \caption{Comparison between IMLS with Gaussian weight (left) and EIMLS (right). The upper row is computed with $h_0=0.0015$, and the lower row is computed with $h_0=0.0001$. The IMLS formulation is not defined over the whole computational domain (white parts of the figure) whereas EIMLS is.}
  \label{fig:eimls}
\end{figure}

In this paragraph, we present the IMLS surface representation. We also show why it needs to be extended to meet the requirements of the proposed framework. All the used notations are gathered in Table~\ref{tab:implicit_notations}.

IMLS have been introduced by Kolluri \cite{kolluri_provably_2005}. He gave an implicit closed form to express the MLS projection operator:

\begin{equation}
  \alpha_{IMLS}(x) = \frac{ \sum_{i\in\mathcal{N}}{ w_i(x) (x - p_i).n_i } }{ \sum_{i\in\mathcal{N}}{ w_i(x) } }
  \label{eq:kol05_imls}
\end{equation}

A point $p_i$ and its normal $n_i$ define a local tangent plane to the surface. The term $(x - p_i).n_i$ is the signed distance from $x$ to this plane. Equation \ref{eq:kol05_imls} can then be interpreted as a weighted sum of the signed distances to the tangent planes to $\mathcal{S}_\mathcal{P}$ at each $p_i$.

$w_i$ is the weighting function associated with the point $i$. The weighting function can be derived from a generic weighting function:

\begin{equation}
  w_i(x) = \phi \Bigg(\frac{\norm{p_i - x}}{h_i(x)} \Bigg)
  \label{eq:weighting_imls}
\end{equation}

$h_i$ is a space parameter, and it is usually chosen to be constant for each point. If the sampling is close to uniform over the whole point cloud, it can be chosen constant for all the points and from the order of magnitude of the sensor noise $h_i=h$. If the point cloud is sparse, the space parameter can be chosen from the order of magnitude of the local density of $\mathcal{P}$. 

The choice of the generic weighting function $\phi$ can yield different behaviors of the underlying surface. Here are some examples:
\begin{itemize}
  \item \textbf{Gaussian weight:} This is the classical weight used by \cite{kolluri_provably_2005}
  \begin{equation}
  \phi(x) = e^{-\frac{x^2}{2}}
  \label{eq:gaussian_weight}
  \end{equation}
  
  With a Gaussian weight, the surface is approximated by smoothing out the noise. As it is not compactly supported, the surface is, in theory, well defined in the whole domain of study. We will see later why we cannot rely on this assumption.
  
  \item \textbf{Compactly supported weight:} As in \cite{guennebaud_algebraic_2007}, the weights can be defined from a compactly supported polynomial function:
  \begin{equation}
  \phi(x) = \left \{ \begin{array}{ll}
      (1-x^2)^4 & \text{if } x<1 \\
      0         & \text{otherwise}
   	\end{array} \right.
  \end{equation}
  
  This weight is compactly supported so the IMLS implicit function is defined only on the union of the open balls of radius $h_i$ and centered on each point. This definition domain is not guaranteed to cover the whole domain of study.
  
  \item \textbf{Interpolatory weight:} The weighting function can also be chosen to interpolate the points if it is chosen with a singularity on zero, as for example:
  \begin{equation*}
  \phi(x) = \frac{1}{x^2}
  \end{equation*}  
  For real-world point clouds, an interpolatory surface is not desirable as it would over-fit the noise.
\end{itemize} 

In the following, we consider only the Gaussian weighting function (given by Equation \ref{eq:gaussian_weight}). It is theoretically defined for all $x\in\mathbb{R}^+$, and thus, $\alpha_{IMLS}$ should be defined in the whole space $\mathbb{R}^d$, but in practice, it is defined only until it falls below the numerical precision.

Let $10^{-\gamma}$ be a numeric limit. We have:

\begin{equation}
  \phi(x) < 10^{-\gamma} \Rightarrow x > \sqrt{ 2 \gamma \text{ log}(10)} = l_\gamma
\end{equation}

This means that the Gaussian weight is, in fact, compactly supported, which means that for a query point sufficiently far from the point cloud, all the weights $w_i(q)$ are null, and thus, $\alpha_{IMLS}$ is not defined. We can see in the left of Figure \ref{fig:eimls} that the definition domain for IMLS computed with the base Gaussian weight expression (a) does not cover the whole domain of study and (b) does not produce a watertight surface for low values of $h$.

This is an issue for our workflow because the implicit function must be defined over the whole domain to drive both the mesh adaptation and the flow solver. In addition, for flow simulation, the domain is generally several times larger than the object extension. Thus, the implicit scalar function must be defined far from the surface. For all these reasons, the classical IMLS formulation is not suited to this problem and must be extended. In the following, we present our novel EIMLS extended formulation.

\subsubsection{Extended implicit moving least squares (EIMLS)}

We first introduce the $k$ nearest neighbor operator on a point cloud.

\begin{defn}
  \textit{The $k$ nearest neighbor (KNN) operator on the point cloud}
  
  Given $x \in \mathbb{R}^d$ and $k$ a positive integer, the set of indices of the $k$ closest points to $x$ in $\mathcal{P}$ is noted:
  \begin{equation}
  \pi_\mathcal{P}(k,x) \in P_k(\mathbb{N})
  \label{eq:knn_operator}
  \end{equation}

  In the special case $k=1$, $\pi_\mathcal{P}(1,x)$ defines the nearest neighbor (NN) projection operator on $\mathcal{P}$. For ease of notation, it is noted:
  \begin{equation}
  \pi_\mathcal{P}(x) \in \mathbb{N}
  \label{eq:nn_operator}
  \end{equation}
\end{defn}

We propose the following extended expression for the space parameter:
\begin{equation}
h^E(x) = \text{max}\Bigg(\frac{ \norm{p_{\pi_\mathcal{P}(x)} - x} }{ l_\gamma } , h_0 \Bigg)
\label{eq:extended_space_param}
\end{equation}

\noindent where $h_0$ is the classical constant space parameter that can be chosen from the order of magnitude of the point cloud noise. EIMLS can then be computed with Equations \ref{eq:kol05_imls}, \ref{eq:weighting_imls}, \ref{eq:gaussian_weight} and \ref{eq:extended_space_param}. Thanks to this definition, we ensure that a least one weight $w_i$ does not fall below the numeric precision threshold: the one associated with $\pi_\mathcal{P}(x)$; the closest point to $x$ in the point cloud.

We can see on the right of Figure~\ref{fig:eimls} that EIMLS yields the same results as IMLS where it is defined. We can also notice that (b) EIMLS is well defined over the whole domain of study and (d) EIMLS is much more robust to small values of $h$ providing a definition extended to the whole domain. One can notice that the value far from the surface is not necessarily exact but it is not an issue for the rest of our workflow as we need only precision near the surface. This point is clarified in Section~\ref{sec:anisotropic_mesh_adaptation}.

\subsection{Implementation}

\paragraph*{Neighbor computations} If we use Equation~\ref{eq:kol05_imls} with a non-compactly supported weight, we should sum the contributions of all the points of $\mathcal{P}$. This is not coherent with the previous claim that IMLS is a local surface representation. However, as a decreasing weight is numerically null when it falls below the numerical precision, it is common to perform the following approximation. Instead of summing over $\mathcal{N}$, we only sum over a $\pi_\mathcal{P}(k,x)$:
\begin{equation}
  \widehat{\alpha}_{EIMLS}(x) = \frac{ \sum_{i \in \pi_\mathcal{P}(k,x)}{ w(x) (x - p_i).n_i } }{ \sum_{i \in \pi_\mathcal{P}(k,x)}{ w(x) } }
  \label{eq:kol05_imls_approx}
\end{equation}

This way, the representation is only local as it relies only on neighborhood computation. Though, care must be taken when choosing the $k$ parameter: Too small values produce erroneous results, and too large values are computationally expensive. It also depends on the local sampling pattern on the point cloud; for an anisotropic sampled point cloud (as it can arise when the scanner sampling is not uniform in all directions), a combination between a radius search and a KNN search can generate better results. We get rid of these problems with the uniform sub-sampling pass operated on the point clouds. In practice, a value of 80 neighbors worked well for all the datasets.

To speed-up the nearest neighbor computation, we use a KD-tree-based search \cite{moore_intoductory_1991}. The complexity is then $\text{log}(\abs{\mathcal{N}})$ instead of $\abs{\mathcal{N}}$ for the naïve algorithm. We use the fast implementation provided by the widely used FLANN library \cite{muja_fast_2009}.

\paragraph*{Memory concerns} Memory is not an important problem for today's computers when computing EIMLS. However, in our case, it is supposed to run in a massively parallel context. Then the memory of each process is independent from the others even if they are physically present on the same computational node. This means that the entire point cloud should be loaded by each process. In our case, this is not an issue as the clouds we use have a relatively small memory footprint. For larger point clouds, we can use a more compact representation based on octree compression \cite{rusinkiewicz_qsplat:_2000} or on more efficient sparse voxel directed acyclic graphs (DAG) \cite{kampe_high_2013}. In addition, to avoid loading the whole point cloud, an out-of-core loading algorithm can be used in conjunction with loading only the needed parts of the dataset, and thus,s reduce the program's memory use.

%% file: 5_anisotropic_mesh_apadtation.tex
\newcommand{\XMetric}{  \sum_{j \in \Gamma(i)}{ \mathbf{X}^{ij} \otimes \mathbf{X}^{ij} }  }

\section{Anisotropic mesh adaptation on point-sampled surfaces}
\label{sec:anisotropic_mesh_adaptation}

\begin{table}
  \centering
  \begin{tabular}{ll}
    \hline
    Notations                         & Definition \\
    \hline
    $d$                               & dimension of space \\
    $\Omega$                          & computational domain \\
    $\mathcal{N} \subset \mathbb{N}$  & set of node indices \\
    $\mathcal{X} = \{\mathbf{X}^i, i \in \mathcal{N} \} \subset \mathbb{R}^d$  & set of nodes \\
    $\mathbf{X}^{ij} = \mathbf{X}^{j} - \mathbf{X}^{i} \in \mathbb{R}^d$ & edge vector between two connected nodes $i$ and $j$ \\
    $\Gamma(i) \subset \mathcal{N}$   & set of nodes connected to node $i$ \\
    $\mathbf{G}^i$                    & reconstruction gradient operator at node $i$ \\
    $\mathbb{M}^i$                    & unit mesh metric at node $i$ \\
    \hline
  \end{tabular}
  \caption{Notations and definitions for Section \ref{sec:anisotropic_mesh_adaptation}}
  \label{tab:mesh_adap}
\end{table}

In this section, we present the mesh adaptation step of the proposed method (step (c) in Figure~\ref{fig:our_method}). This step is used to create a volumetric and anisotropic unstructured tetrahedral grid, which is adapted around the implicit surface computed by EIMLS directly on a point cloud. The anisotropic adaptation is performed to minimize an \textit{a posteriori} error metric \cite{coupez_metric_2011}. The notations used in this section are shown in Table~\ref{tab:mesh_adap}.

\subsection{Related work}

We use an immersed boundary finite element Navier-Stokes flow solver to compute the flow around the point-sampled geometry. As explained by \cite{mittal_immersed_2005}, the immersed boundary method has been introduced by \cite{peskin_flow_1972}. This method allows the simulation to be performed on a grid that fills the entire simulation domain instead of only the flow part of the domain, in the classic body-fitted mesh paradigm. It then suppresses all the complex geometry-specific problems in the mesh construction. This method has been initially developed to run on Cartesian grids, but more recently, it has been applied to anisotropic unstructured grids. An anisotropic unstructured grid provides better numerical accuracy for the same computational nodes budget but requires a more sophisticated meshing algorithm.

We use a local optimization meshing algorithm \cite{coupez_generation_2000, coupez_edge-based_2013}. As it is based only on local mesh modifications of an existing mesh, it can be used for both meshing and remeshing and is massively parallel \cite{coupez_parallel_2000}. Compared to traditional meshing algorithms, this algorithm is much less computationally intensive for remeshing as the mesh is not rebuilt from scratch. This way, the mesh can be adapted around both the geometry of the immersed boundary and the flow itself between each iteration of the transient flow solver.

\subsection{Truncated signed distance function (TSDF) for mesh adaptation}

Let $\Omega$ be the simulation domain and $\omega_\mathcal{P}$ the sub-domain of $\Omega$ delimited by the point-sampled boundary $\mathcal{S}_\mathcal{P}=\delta\omega_\mathcal{P}$. The EIMLS defined in the previous section approximates the signed distance function to the surface. Let $D(x,\mathcal{S}_\mathcal{P})$ be the unsigned distance function to $\mathcal{S}_\mathcal{P}$. The EIMLS implicit function verifies near the surface:

\begin{equation}
  \alpha_{EIMLS}(x) \approx \left \{
  \begin{array}{rl}
    -D(x,\mathcal{S}_\mathcal{P}) & \mbox{if } x \in \omega_\mathcal{P} \\
     D(x,\mathcal{S}_\mathcal{P}) & \mbox{else}
  \end{array} \right.
\end{equation}

If $\mathcal{S}_\mathcal{P}$ is a smooth surface, $\alpha_{EIMLS}$ is then differentiable almost everywhere, and $\norm{\nabla{\alpha_{EIMLS}}} = 1$. In the following, we consider a $P^1$ mesh approximation of the $\alpha_{EIMLS}$ function. The interpolation error along each edge of the mesh is then of order $2$. For this reason, we introduce a strong variation of the second derivative of the implicit function near the surface by using an auxiliary hyperbolic tangent function, as described in \cite{coupez_implicit_2015}:

\begin{figure}
  \centering
  \includegraphics[width=0.7\textwidth]{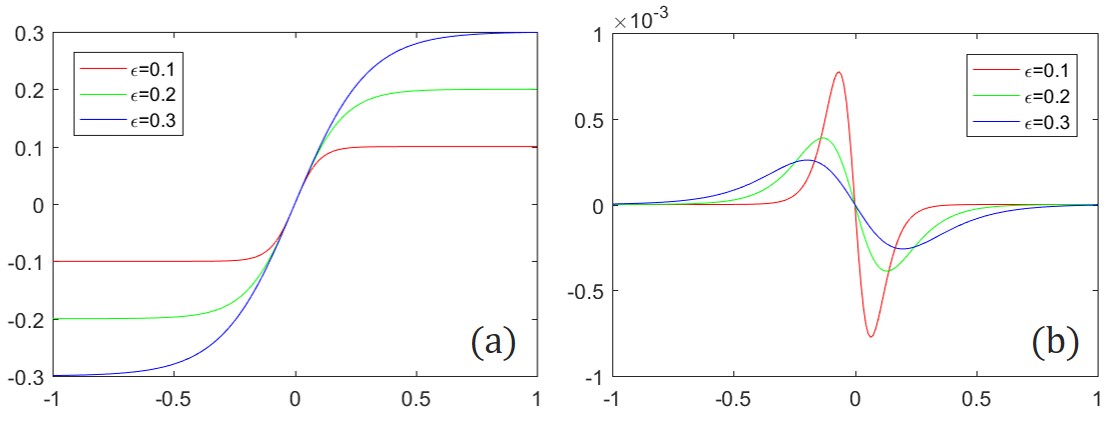}
  \caption{(a) $f_\epsilon$ function. (b) Its second derivative for various values of $\epsilon$. (red) $\epsilon=0.1$. (green) $\epsilon=0.2$. (blue) $\epsilon=0.3$.}
  \label{fig:hyperbolic_tangent}
\end{figure}

\begin{equation}
  f_\epsilon(x) = \epsilon \cdot \tanh\left(\frac{x}{\epsilon}\right)
\end{equation}

As shown in Figure~\ref{fig:hyperbolic_tangent}, $f_\epsilon$ is designed to keep a derivative equal to $1$ around $x=0$ for any value of $\epsilon$. Its second derivative is sharper for small values of $\epsilon$. Applying it to $\alpha_{EIMLS}$ yields a smooth truncated signed distance function:

\begin{equation}
  \alpha_\epsilon(x) = f_\epsilon \left(\alpha_{EIMLS}(x)\right)
\end{equation}

We can see the effect of $\epsilon$ on $\alpha_\epsilon(x)$ for various values of $\epsilon$ in the last row of images in Figure~\ref{fig:adap_epsilon}. This figure also shows the effect of $\epsilon$ on the mesh adaptation algorithm which is discussed later.

As we use only a truncated signed distance function, numerical accuracy is not needed far from the surface. This is why EIMLS is well suited for this problem as it may not be numerically exact far from the surface if we compare it to the true signed distance to the surface, but it is sufficient as it is correct near the surface and keeps the sign coherence.

\subsection{A posteriori interpolation error estimation}

\cite{coupez_metric_2011} introduced an \textit{a posteriori} error estimate based on the length distribution tensor and edge-based error analysis. \cite{coupez_edge-based_2013} then extended this work to constrain the overall node number in the resulting adapted mesh. The reader can refer to these works for detailed demonstrations of the presented results.

Let $u\in\mathcal{C}^2(\Omega)$ a scalar field known at the mesh nodes $\mathcal{X} = \{\mathbf{X}^i \in \mathbb{R}^d, i \in \mathcal{N}\}$. We note: $U^i = u(\mathbf{X}^i)$ and $U^{ij} = U^j - U^i$. \cite{coupez_metric_2011} show that the gradient interpolation operator $\mathbf{G}^i$ is defined by:

\begin{equation}
  \mathbf{G}^i = \left(\XMetric\right)^{-1} \sum_{j \in \Gamma(i)}{ U^{ij} \mathbf{X}^{ij}  }
\end{equation}

This operator can then be used to define $e_{ij}$ the approximation interpolation error along each edge:

\begin{equation}
  e_{ij} = \abs{ \mathbf{G}^{ij} . \mathbf{X}^{ij} }
\end{equation}

Given a mesh, we can construct a unit metric field $\{ \mathbb{M}^i \}_{i \in \mathcal{N}}$ that represents the statistical edge distribution at each node:

\begin{equation}
  \mathbb{M}_i = \frac{\abs{\Gamma(i)}}{d} \left(\XMetric\right)^{-1}
\end{equation}

This metric field is a representation of the mesh local elements shape and size. Conversely, from a mesh metric field, we can build a mesh whose elements conform to the local shape and size defined by the metric field. We use this last observation to define a new metric field $\widetilde{\mathbb{M}}^i$ from a stretching transformation applied to all edges of the mesh. Let's then suppose that all edges are stretched by a stretching coefficient $s_{ij} \in \mathbb{R}^+$:

\begin{equation}
  \left\{
  \begin{array}{l}
  \widetilde{\mathbf{X}}^{ij} = s_{ij} \mathbf{X}^{ij} \\
  \widetilde{e}_{ij} = s_{ij}^2 e_{ij} \\
  \end{array}
  \right.
\end{equation}

We then define $n_{ij}$ as the number of created edges along each edge $\{i,j\}$:

\begin{equation}
  n_{ij} = s_{ij}^{-1} = \sqrt{ \frac{e_{ij}}{\widetilde{e}_{ij}} }
\end{equation}

We then suppose that the new edge error is constant over the whole mesh $\widetilde{e}_{ij} = e$. This way, the error is balanced such that the over-detailed regions are coarsened, and the under-detailed ones are refined in an anisotropic fashion. However, the problem is still ill posed as this error also depends on the resulting number of nodes. We then suppose also that the overall number of nodes is constant and equal to $N$.

\cite{coupez_edge-based_2013} show that the number of nodes created per node $i$ is given by:

\begin{equation}
    n^i(e) = e^{-d/2}  \underbrace{ \det \left(
        \left( \sum_{j \in \Gamma(i)}{
        \frac{\mathbf{X}^{ij}}{\norm{\mathbf{X}^{ij}}} \otimes \frac{\mathbf{X}^{ij}}{\norm{\mathbf{X}^{ij}}} } \right)^{-1}
        \left( \sum_{j \in \Gamma(i)}{ \sqrt{e_{ij}} \frac{\mathbf{X}^{ij}}{\norm{\mathbf{X}^{ij}}} \otimes \frac{\mathbf{X}^{ij}}{\norm{\mathbf{X}^{ij}}} } \right)
        \right) }_{n^i(1)}
\end{equation}

Thus, the total number of nodes in the resulting mesh is given by:

\begin{equation}
  N = e^{-d/2}  \sum_i n_i(1)
\end{equation}

Then we can express the overall error $e$ in function of $N$:

\begin{equation}
  e = \left( \frac{\sum_i n_i(1)}{N} \right)^{2/d}
\end{equation}

The new metric corresponding to a balanced error over the whole domain is given by:

\begin{equation}
  \mathbb{\widetilde{M}}_i(e) = \frac{1}{e} \frac{\abs{\Gamma(i)}}{d} \left(  \sum_{j \in \Gamma(i)}{ \frac{1}{e_{ij}} \mathbf{X}^{ij} \otimes \mathbf{X}^{ij} }  \right)^{-1}
\end{equation}

\subsection{Anisotropic mesh adaptation on point-sampled surface}
\label{subsec:anisotropic_mesh_adaptation_on_point_sampled_surfaces}

\begin{figure}
  \centering
  \includegraphics[width=\textwidth]{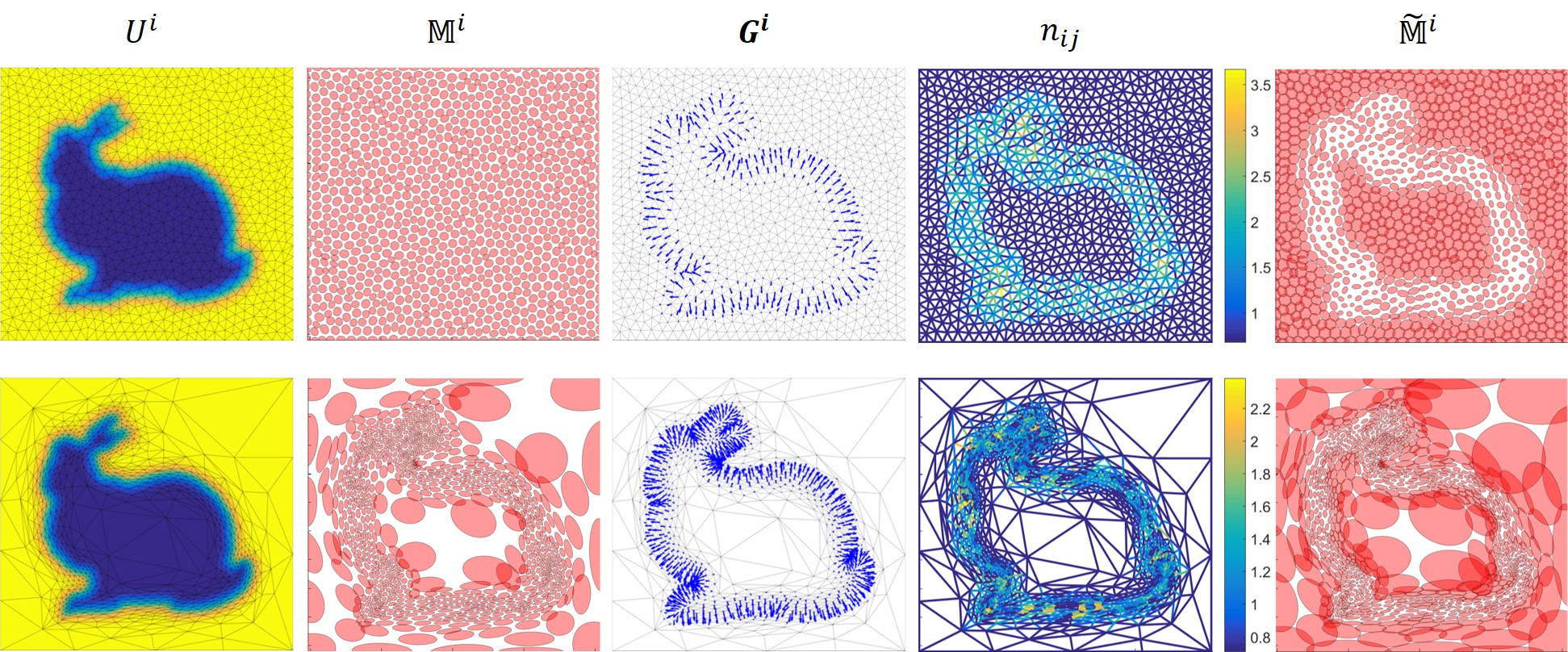}
  \caption{Adapted mesh metric computation workflow. (top row) Starting with an isotropic mesh. (bottom row) Starting with an already adapted mesh. See subsection~\ref{subsec:anisotropic_mesh_adaptation_on_point_sampled_surfaces} for more details.} 
  \label{fig:adap_workflow}
\end{figure}

Figure~\ref{fig:adap_workflow} shows the metric construction steps described in the previous subsection. The upper row corresponds to a uniform grid and the lower to an already adapted mesh. The same truncated EIMLS scalar function is sampled over the two meshes with $h_0=0.003$ and $\epsilon=0.005$. The unit mesh metric $\mathbb{M}^i$ is represented in the second row with ellipsoids centered on each node. Each ellipsoid is the set of points at a distance (with respect to the node metric) less than or equal to $1$ from the node. Each ellipsoid is shown at 50\% scale for better visual understanding. This representation shows clearly that the unit metric field describes the shape and size of the mesh elements. The third column depicts the reconstruction gradient operator $\mathbf{G}^i$. Its variation perpendicular to the surface is steeper over the uniform mesh than over the adapted mesh. The number of created nodes per edge $n_{ij}$ (drawn in the fourth column) is then high along this direction over the uniform mesh, whereas it is almost constant near the surface for the adapted mesh. The resulting metric $\widetilde{\mathbb{M}}^i$, represented in the fifth column, is then anisotropic near the surface for the uniform mesh. The resulting mesh (after a subsequent remeshing step) would effectively be coarsened far from the surface and refined anisotropically near the surface. The resulting metric for the adapted mesh is very similar to the initial mesh unit metric $\mathbb{M}^i$ which means that the mesh is already well adapted over the domain.

%% file: 6_results.tex
\section{Results}
\label{sec:results}

In this section, we demonstrate the ability of the proposed method to compute flows around point-sampled geometries. First, we show results for the mesh adaptation around the surface. We then present external aerodynamics simulations on the 2D dataset and the three 3D datasets described in Figure~\ref{fig:datasets_3d}. All the results have been produced with the ICI-lib code developed at the High Performance Computing Institute (ICI). The code implements an anisotropic unstructured mesher by local optimization and a Navier Stokes transient finite element immersed boundary flow solver \cite{coupez_solution_2013}.

\subsection{Mesh adaptation: geometry reconstruction}
We first demonstrate the ability of our method to capture an implicit surface defined on a point cloud. Capturing the surface well is a critical issue as it defines how well the geometric details are represented and thus, how well the flow solver is able to handle these details.

\paragraph*{2D iterative adaptive anisotropic meshing}

\begin{figure}
  \centering
  \includegraphics[width=\textwidth]{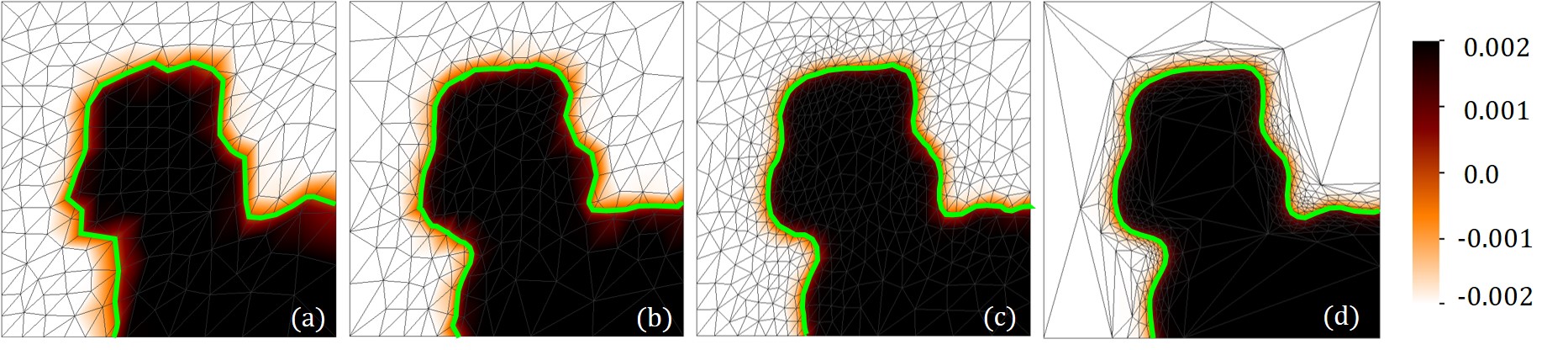}
  \caption{Illustration of the ability of the proposed method to capture iteratively a surface. The zero level set of the implicit function interpolated on the mesh is depicted in green. The scale in the right lower corner gives the value of the truncated EIMLS function ($h_0=0.003$ and $\epsilon=0.002$). (a) Initial isotropic mesh. (b) After 10 remeshing iterations. (c) After 15 remeshing iterations. (d) After 30 remeshing iterations.}
  \label{fig:results_adap_2d}
\end{figure}

Figure~\ref{fig:results_adap_2d} shows how the proposed method captures iteratively 2D point-sampled geometry with a constant number of nodes. The initial mesh is isotropic, and the interpolation error is big near the surface due to the truncated signed implicit function. The meshing algorithm is then performed iteratively switching between metric computation and remeshing. The implicit function is evaluated at the mesh nodes only once per iteration. The metric is also computed once per iteration. During an iteration, when new nodes are created the implicit function and the metric are interpolated at the new node location.

\paragraph*{Influence of $\epsilon$}

\begin{figure}
  \centering
  \includegraphics[width=0.8\textwidth]{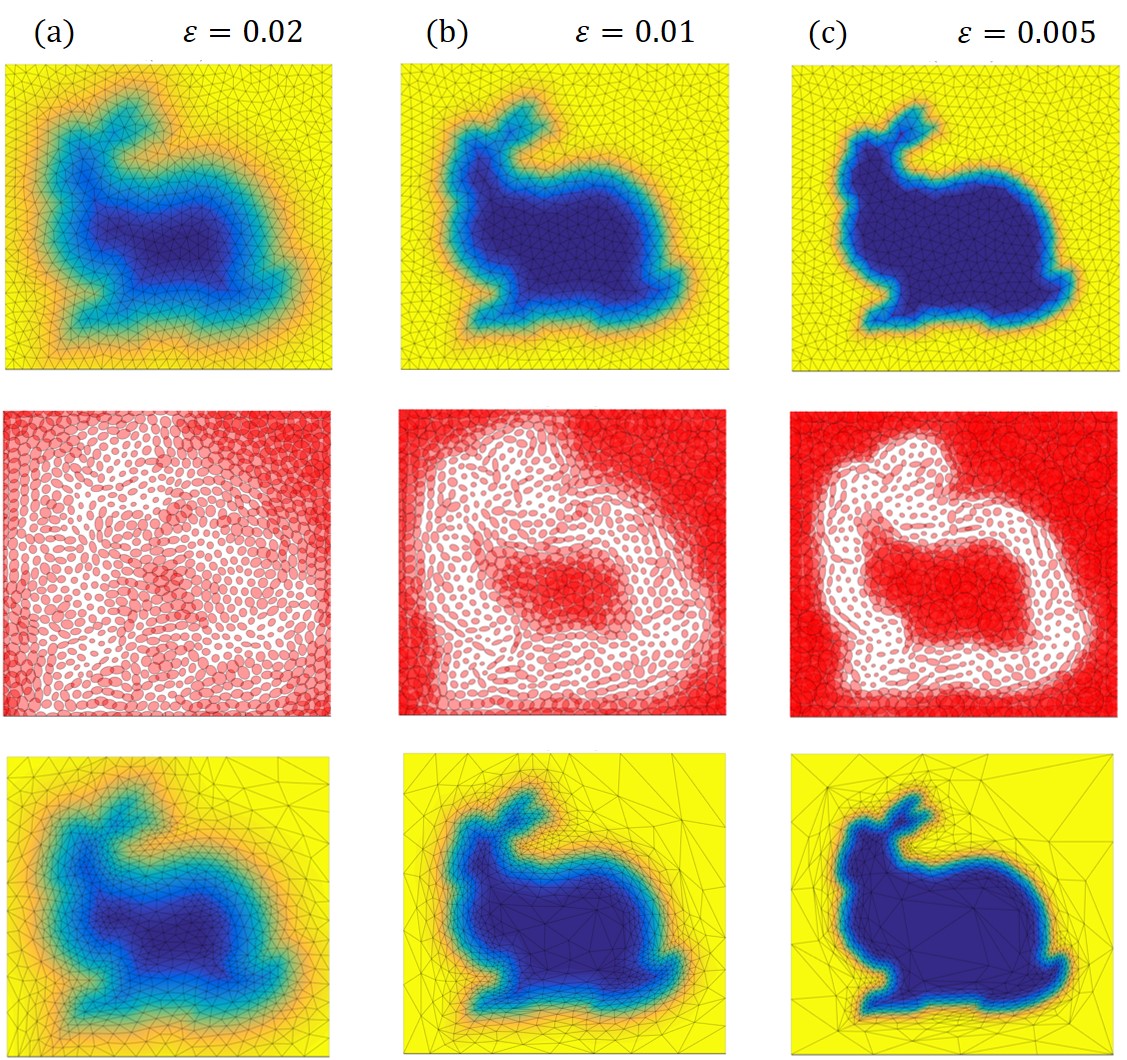}
  \caption{Influence of the $\epsilon$ parameter of the hyperbolic tangent function. (top row) Implicit EIMLS function sampled on the mesh nodes and interpolated over the mesh. (central row) New metric defined with the edge distribution tensor and per-edge error. (bottom row) Mesh adapted conform to the new metric. The metric is represented here with ellipsoids centered on the mesh nodes. For each node, the ellipsoids contain all the points at a distance of 0.5 with respect to the metric.} 
  \label{fig:adap_epsilon}
\end{figure}

The $\epsilon$ parameter introduced in the previous section determines how steep the gradient variation is near the surface. We illustrate in 2D in Figure~\ref{fig:adap_epsilon} the influence of this parameter. Large values of $\epsilon$ (a) compared to the object scale produce a constant and isotropic metric field computed from the edge error on the derivative of the gradient of the scalar function interpolated on the mesh. The mesh is then less adapted than for smaller values of $\epsilon$ (b) and (c). In addition, for smaller values of $\epsilon$ (c) near the surface, the metric is anisotropic, and more elongated in the tangent direction from the surface.

\paragraph*{3D adaptive anisotropic meshing}

\begin{figure}
  \centering
  \includegraphics[width=\textwidth]{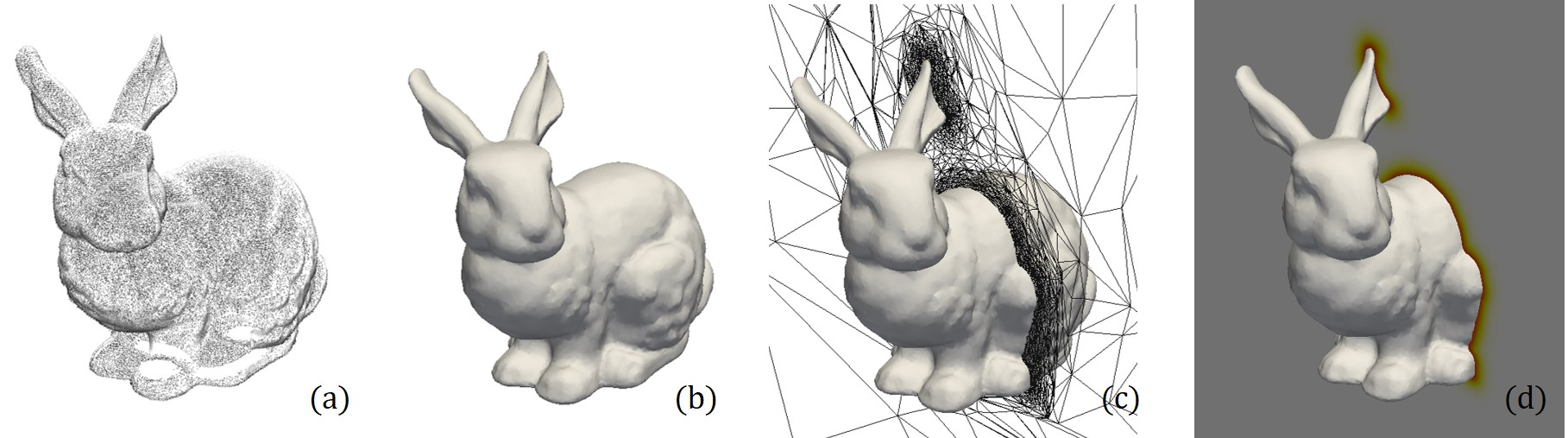}
  \caption{Mesh adaptation around the 3D implicit function defined from a point cloud. (a) Initial point cloud: Stanford Bunny (362,270 points). (b) Zero level extracted from the implicit EIMLS function interpolated on the final adapted mesh. (c) Slice of the adapted mesh. (d) Slice of the implicit function.}  
  \label{fig:results_adap_3d}
\end{figure}

Figure~\ref{fig:results_adap_3d} depicts the results of the proposed method for mesh adaptation on 3D point-sampled surfaces. The extracted iso-surface from the interpolated implicit function on the mesh is identical to the original point-sampled model. All the details on the surface are well preserved, and the holes in the model are filled.

\subsection{Flow simulation: test cases}

In this subsection, we show various examples of flow simulation performed with the proposed method on complex point-sampled geometries.

\subsubsection{Dynamic multi-criteria mesh adaptation}

\begin{figure}
  \centering
  \includegraphics[width=0.4\textwidth]{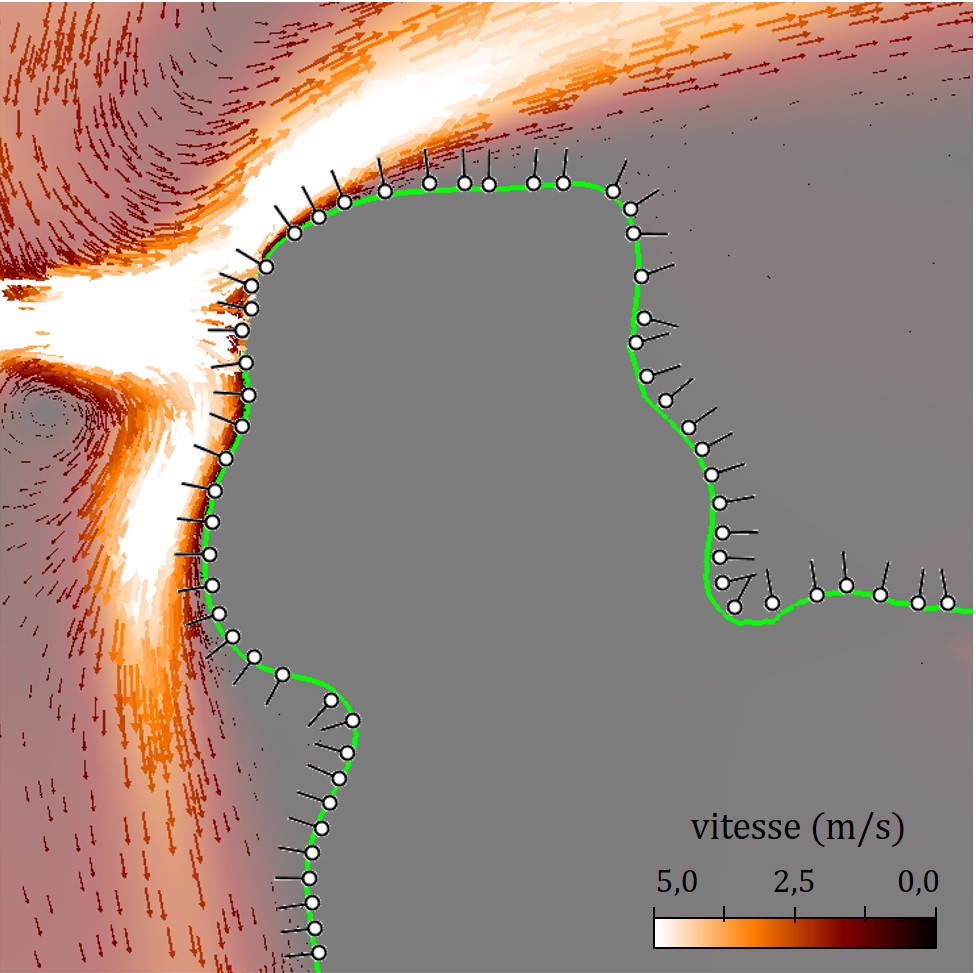}
  \includegraphics[width=0.4\textwidth]{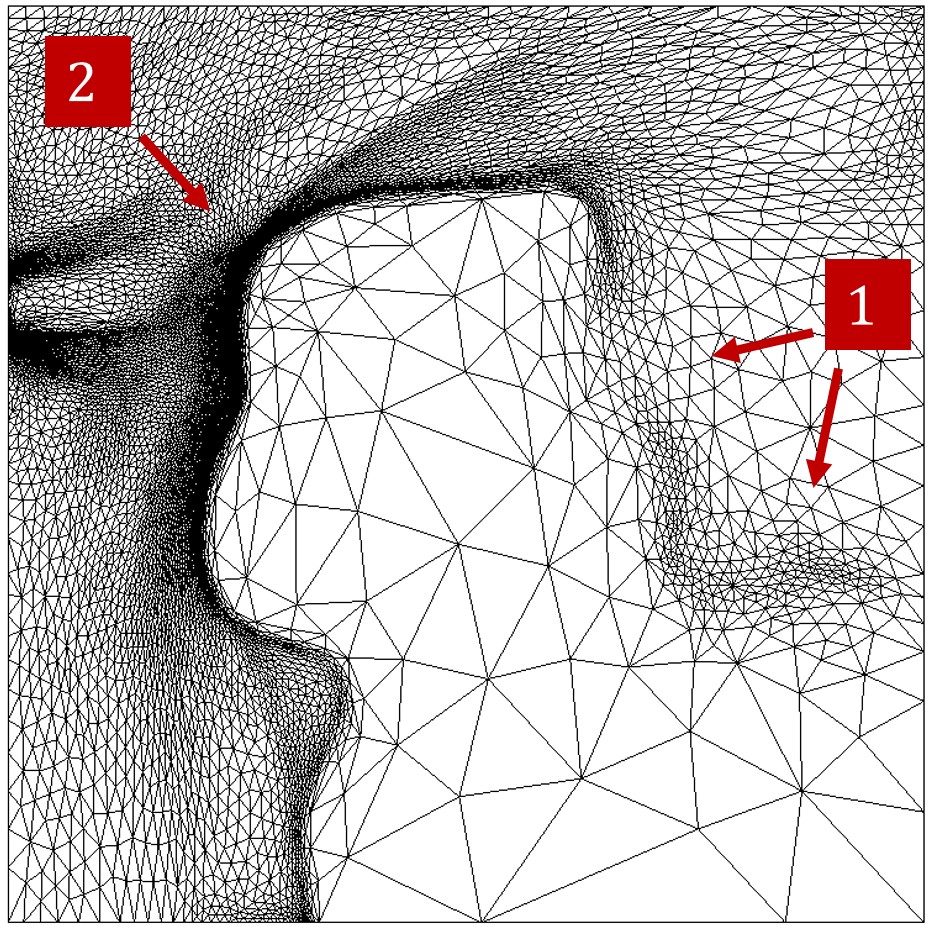}
  \caption{(left) The flow around the 2D point cloud. Air is blown from a small inlet on the left of the domain at 5 m/s. The flow splits in two when it reaches the surface. (right) The mesh adapted around the surface of the obstacle. (1) This is a portion of space where the flow is calm and then geometric accuracy is not needed, so the mesher puts fewer computational nodes here. (2) In contrast, here the mesh resolution is very high because it is adapted around the surface and around high-gradient regions of the flow.}
  \label{fig:results_flow_2d}
\end{figure}

Figure~\ref{fig:results_flow_2d} shows a multi-criteria mesh adaptation in 2D. The mesh is adapted around the surface and the flow. This is performed at the same time and requires no \textit{a priori} knowledge of the flow behavior. The mesh adaptation used here gives better control on the computational nodes on the domain. For example, in some parts of the domain where the flow velocity is slow, the mesher puts fewer computational nodes because out there numerical precision on the geometry is less important than in other parts of the domain. However, the underlying geometric precision is still high because it is carried by the detailed point cloud.

\subsubsection{2D and 3D flow simulation}

\begin{figure}
  \centering
  \includegraphics[width=0.7\textwidth]{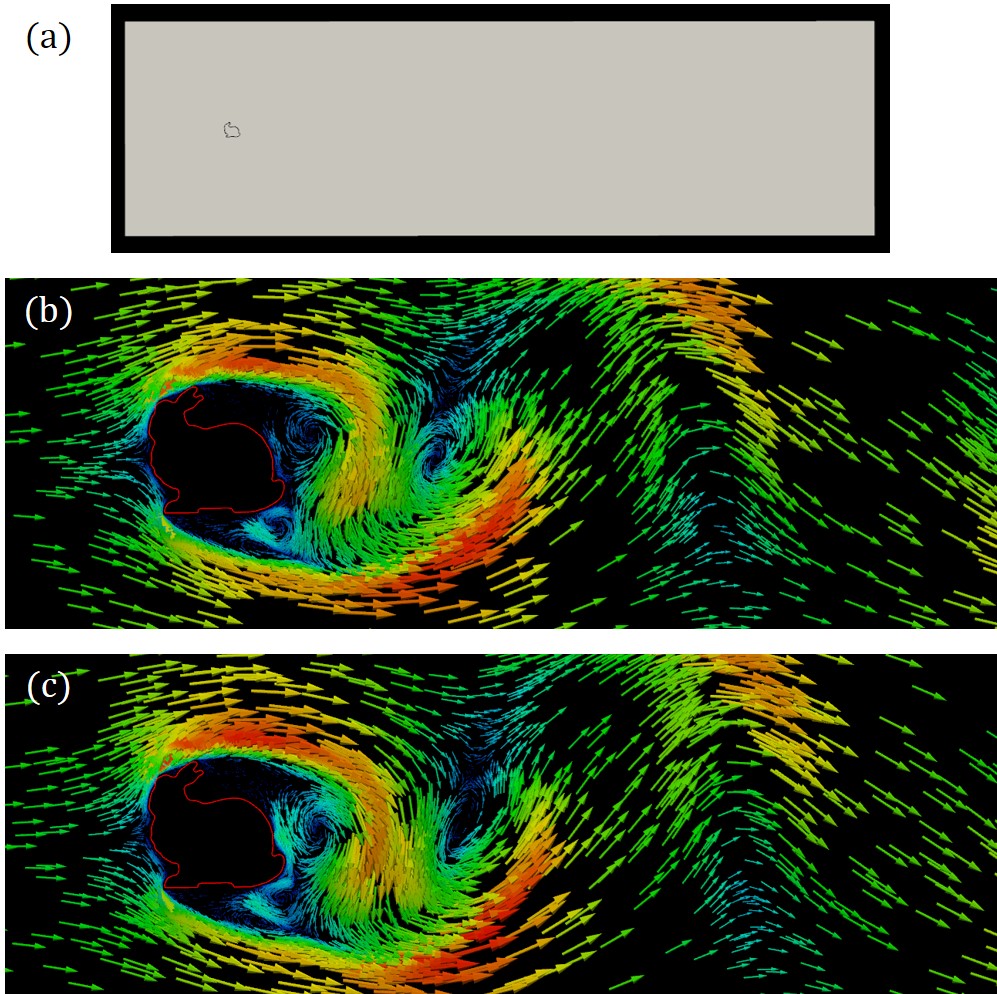}
  \caption{Flow simulation for the 2D slice of the \textbf{bunny} dataset. (a) Computational domain. (b) and (c) Velocity field at different time-steps of the transient simulation.}
  \label{fig:results_bunny_2d}
\end{figure}

\begin{figure}
  \centering
  \includegraphics[width=0.7\textwidth]{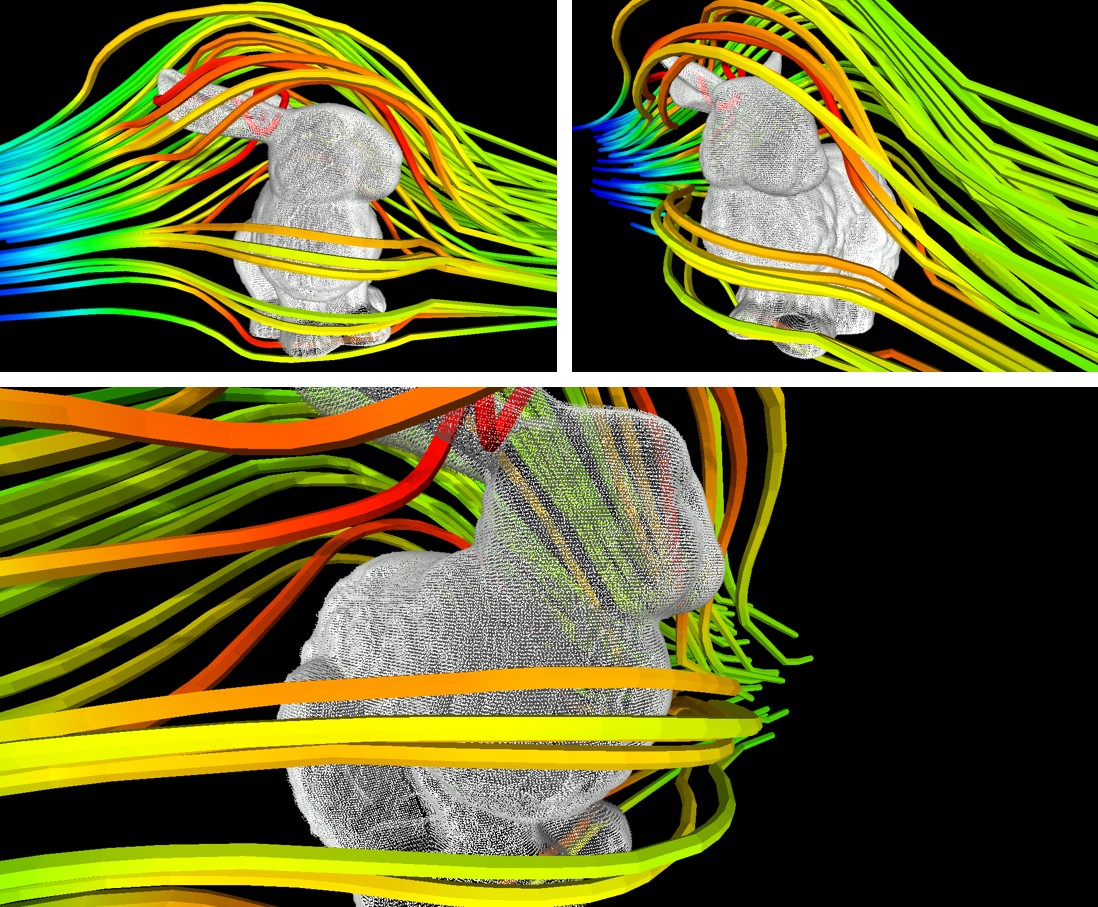}
  \caption{Streamlines around the \textbf{bunny} dataset.}  
  \label{fig:results_flow_3d}
\end{figure}

\begin{figure}
  \centering
  \includegraphics[width=\textwidth]{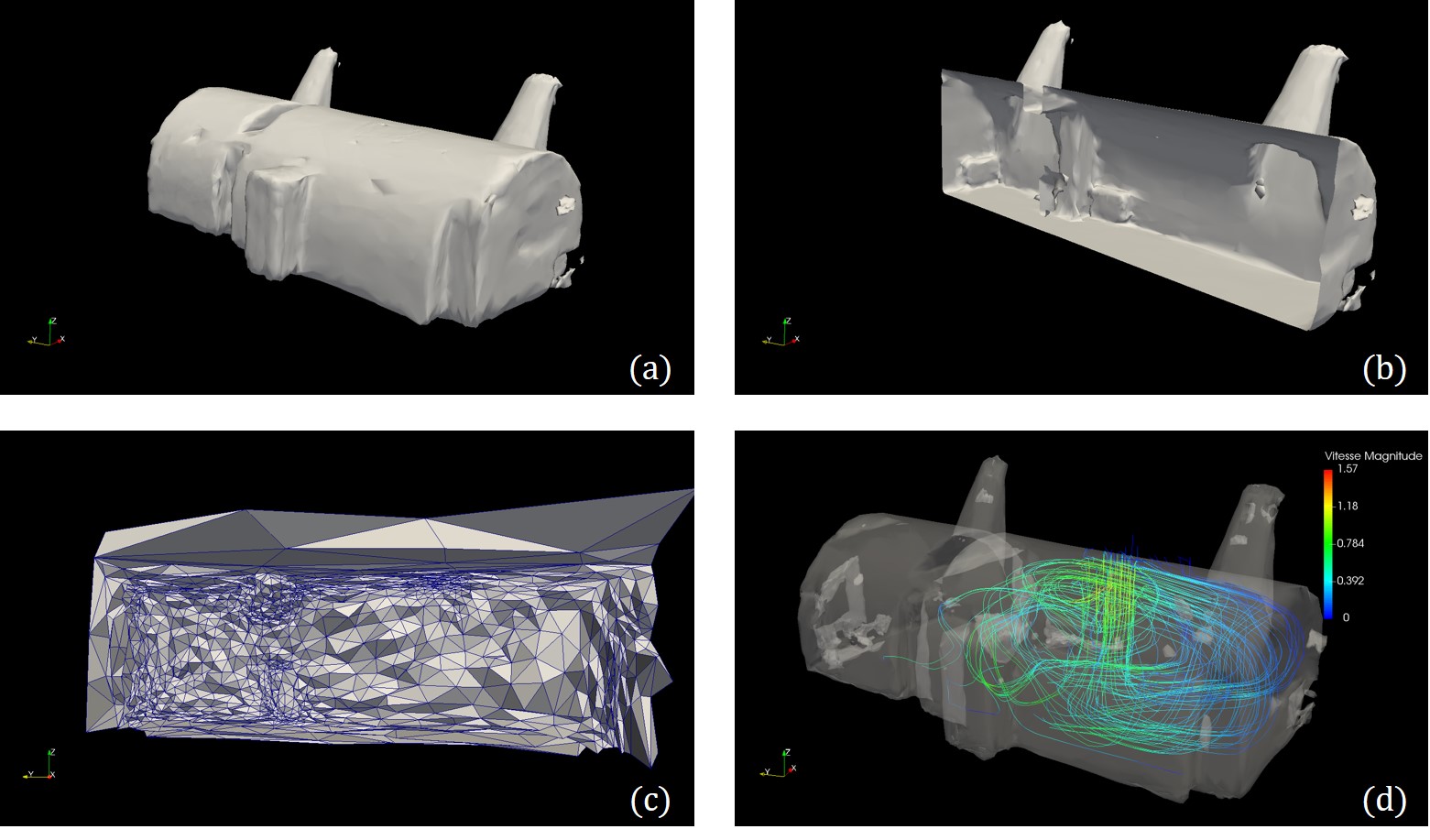}
  \caption{\textbf{Room} dataset. (a) Iso-surface extracted from the EIMLS function sampled on the adapted mesh. (b) Cut of the iso-surface showing the details inside the room. (c) Cut of the adapted mesh. (d) Simulation ventilation in the ceiling of the room.}
  \label{fig:salle_saint_jacques_flow}
\end{figure}

\begin{figure}
  \centering
  \includegraphics[width=\textwidth]{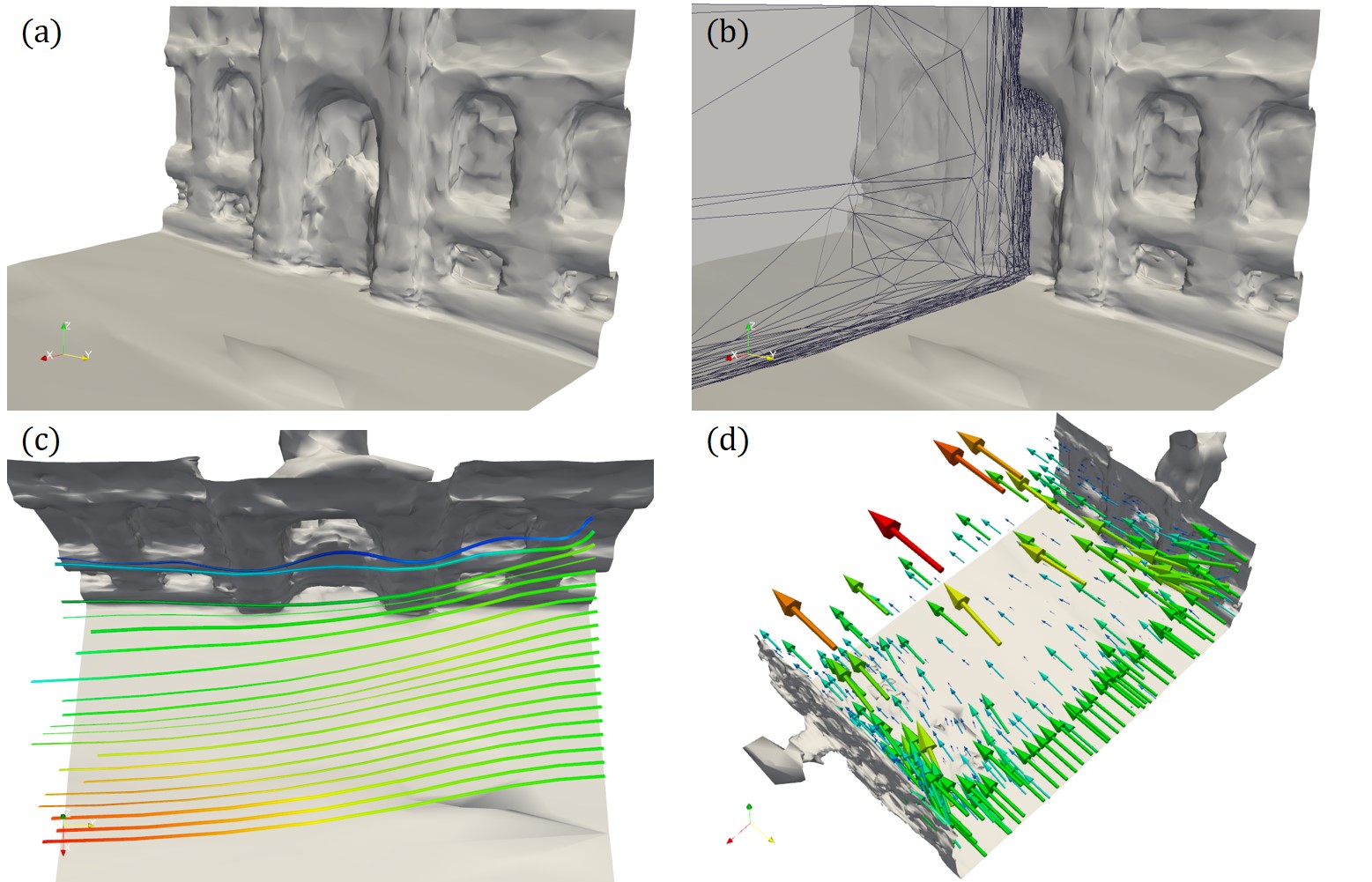}
  \caption{\textbf{Street} dataset. (a) Iso-surface adapted around a gate in the street. (b) Cut of the adapted mesh. (c) streamlines near the gate. (d) Velocity field of the flow.}
  \label{fig:rue_madame_flow}
\end{figure}

Figure~\ref{fig:results_bunny_2d} is another example of simulation on a 2D point cloud. Figure~\ref{fig:results_bunny_2d}(a) illustrates how big the computational domain is compared to the object. The test section is 7 m $\times$ 2 m, whereas the model is only 20 cm tall. The EIMLS surface representation guarantees that the implicit function is well defined over the whole domain. Figure~\ref{fig:results_bunny_2d} also shows the Von Karman vortex street produced by the model. This observed phenomenon is coherent with the flow Reynold number of 200.

Figure~\ref{fig:results_flow_3d} is an example of external aerodynamics on a hand-scanned object. The \textbf{bunny} point cloud is put on a 1 m/s flow. The test section is 5 m long, and its cross section is 1 m $\times$ 1 m. The viscosity is chosen to reach a Reynolds number of 2000. This value is coherent with the observed turbulent flow. We can observe flow recirculation near the bunny ears. The simulation has been performed with 1,000,000 computational nodes in parallel on 100 cores. We used a time-step of 0.0001 s, and it took 3 hours. This type of simulation can be extended to industrial objects, for example, for retro-engineering purposes.

Figure~\ref{fig:salle_saint_jacques_flow} is an example of a ventilation simulation performed on the \textbf{room} dataset, acquired with a terrestrial laser scanner. The fluid is blown from a 20 cm square inlet in the room ceiling. The room has two transverse walls at one third of its length. These walls seem to partition the flow in two. This kind of simulation could be used, for example, to evaluate various ventilation solutions for an existing room or building, based on a highly detailed 3D scan. This could also open up applications in studying the ventilation properties of cultural heritage buildings.

Figure~\ref{fig:rue_madame_flow} is an example of an air ventilation assessment simulation performed at the scale of a mobile 3D laser scanned street. This is relatively new as usually these simulations are performed at the neighborhood or city scale. Simulating air flow at the street level can be used to simulate with a very fine grain of detail the effect of natural disasters or pollution, for example.

%% file: 7_conclusion.tex
\section{Conclusion}

We introduced a new method for simulating flows around 3D point clouds acquired from real-world 3D scans. We demonstrated the ability of the proposed method to capture complex phenomena on various detailed 2D and 3D datasets. We showed the method's ability to use only implicit representation to describe the surface: No explicit mesh reconstruction from the point cloud is needed which reduces the amount of user-interaction and provides the ability to dynamically adapt the mesh during the simulation while keeping all the underlying geometric precision. We also introduced a new EIMLS formulation to define an implicit surface from a point cloud. We demonstrated that this method was able to overcome traditional local implicit surface definitions problems, and well represents complex geometries. Finally, with the development of 3D scanning devices we believe that the proposed method will give the possibility to easily simulate flow on real-life objects opening up both new industrial and research applications.

%% file: aknowledgement.tex
\subsection*{Acknowledgment}

The authors thank {Stanford Computer Graphics Laboratory for the Stanford Bunny 3D scan.

This work was performed by using HPC resources of the Centrale Nantes Supercomputing Centre on the cluster Liger, granted by the High Performance Computing Institute (ICI).